\documentclass[12pt]{article}

\usepackage{dcolumn}
\usepackage{bm}
\usepackage{amsbsy}		
\usepackage{amsfonts}		
\usepackage{amsmath}		
\usepackage{amssymb}		
\usepackage{slashed}
\usepackage{cite}
\usepackage{url}

\usepackage{psfrag}
\usepackage{subfigure}
\usepackage[latin1]{inputenc}
\usepackage[dvips]{graphicx} 
\usepackage{slashed}

\newcommand{\BR}[2]{\mathrm{BR}\mathinner{(#1\rightarrow #2)}}

\newcommand{\mg}{m_{\tilde{g}}}

\newcommand*{\Ave}[1]{\mathinner{\left\langle{#1}\right\rangle}}
\newcommand{\cp}{\mathcal{CP}}

\begin{document}
\begin{center}
\Large\bf\boldmath
\vspace*{0.8cm} Charged Higgs bosons in Minimal Supersymmetry: \\
Updated constraints and experimental prospects
\unboldmath
\end{center}
\vspace{0.6cm}
\begin{center}
D. Eriksson\footnote{Electronic address: \tt david.eriksson@physics.uu.se}, F. Mahmoudi\footnote{Electronic address: \tt nazila.mahmoudi@tsl.uu.se} and O. St\aa l\footnote{Electronic address: \tt oscar.stal@physics.uu.se} \\[0.4cm]
\vspace{0.6cm}
{\sl High-Energy Physics, Dept.~of Physics and Astronomy, Uppsala University,\\ P.\,O.\,Box 535, SE-751\,21 Uppsala, Sweden}
\end{center}

\vspace{0.4cm}
\begin{abstract}
\noindent 
We discuss the phenomenology of charged Higgs bosons in the MSSM with minimal flavor violation. In addition to the constrained MSSM (CMSSM) with universal soft supersymmetry breaking mass parameters at the GUT scale, we explore non-universal Higgs mass models (NUHM) where this universality condition is relaxed. To identify the allowed parameter space regions, we apply constraints from direct searches, low energy observables, and cosmology. We find that values of the charged Higgs mass as low as $m_{H^+}\simeq~135$~GeV can be accommodated in the NUHM models, but that several flavor physics observables disfavor large $H^+$ contributions, associated with high $\tan\beta$, quite independently of MSSM scenario. We confront the constrained scenarios with the discovery potentials reported by ATLAS and CMS, and find that the current exclusion by indirect constraints is similar to the expected LHC discovery reach with $30$ fb$^{-1}$ of data. Finally, we evaluate the sensitivity of the presented discovery potential to the choice of MSSM benchmark scenario. This sensitivity is found to be higher in the case of a light ($m_{H^+}<m_t$) charged Higgs.
\end{abstract}

\section{Introduction}
Charged Higgs bosons are attractive ingredients in theories which extend the Standard Model (SM) Higgs sector with additional fields in non-singlet representations of $\mathrm{SU}(2)_L$. In particular, the observation of a fundamental charged scalar can provide unambiguous evidence of an extended Higgs sector in unfavorable cases when observing a neutral scalar alone would not be sufficient. This could be the case e.g.~in the Minimal Supersymmetric Standard Model (MSSM), with the lightest $\mathcal{CP}$-even Higgs boson having similar properties to the SM Higgs boson, and the heavier neutrals escaping detection. 

Apart from the interest to high energy collider experiments, the charged Higgs plays an important role for several experiments at lower energies. As another mediator of charged current interactions, it can contribute (constructively or destructively) to decay processes without requiring flavor structure beyond the CKM framework. This coupling to flavor physics makes the charged Higgs particularly useful for constraining indirectly the structure of a Higgs sector beyond the SM \cite{Grant:1994ak,Cheung:2003pw,WahabElKaffas:2007xd}.

Awaiting the experimental verdict, which hopefully will be delivered by the LHC, some questions can be addressed. First of all, which models with a charged Higgs have already been probed by existing experiments? Second, what do these experiments imply about the prospects of discovering charged Higgs bosons at the LHC? We discuss these questions in the context of supersymmetry (SUSY) with a minimal extension of the SM Higgs sector. Earlier work along these lines was presented in \cite{Carena:2006ai,Ellis:2007ss,Barenboim:2007sk}.

The outline of this paper is as follows. In section~\ref{sect:2HDM}, the Two Higgs Doublet Model (2HDM) of the MSSM is discussed in more detail, as we review briefly the theoretical aspects of Higgs bosons in the $\cp$-conserving MSSM with minimal flavor violation (MFV). For the numerical studies, we have limited ourselves to two particular classes of MSSM models: the constrained MSSM (CMSSM), and models with non-universal Higgs masses (NUHM). These are introduced in section~\ref{sect:susymodels}.

To the CMSSM and NUHM models we apply constraints from direct searches, low energy observables, and cosmology in order to identify the parameter regions still open for charged Higgs bosons. The constraints are presented in section~\ref{sect:constraints}, including numerical results and uncertainties which are discussed in some detail. This section extends the recent work \cite{Barenboim:2007sk} by including additional constraints, discussing also the heavy charged Higgs, and by allowing larger non-universality in the NUHM Higgs mass parameters.

In section~\ref{sect:experiment} we analyze the impact of applying the constraints by comparing to the collider reach for discovering charged Higgs at the LHC. We consider both the case with $m_{H^+}<m_t-m_b$, which is of particular interest for early LHC running due to the large number of top quarks that will be produced, and $m_{H^+}>m_t$ which becomes interesting at a later stage. Our findings are exemplified on simulation results from ATLAS and CMS and we compare to the current experimental limit from the Tevatron. We also discuss the dependence of the presented experimental reach on the choice of MSSM scenario.  Section~\ref{sect:conclusions} contains a summary and the conclusions.

\section{Charged Higgs Bosons in the MSSM}
\label{sect:2HDM}
To discuss the 2HDM of the MSSM, we introduce the two $SU(2)_L$ doublets
\begin{equation}
H_1=
\left(
\begin{array}{c}
H_1^{0*}\\
-H_1^-
\end{array}
\right)
\quad
 H_2=
\left(
\begin{array}{c}
H_2^{+}\\
H_2^0
\end{array}
\right)
\end{equation}
with hypercharges $Y=\mp  1$. To avoid flavor changing neutral currents (FCNC) at tree level, a discrete symmetry is imposed on the Higgs sector which makes $H_1$ couple exclusively to down type fermions, while $H_2$ couples only to up type fermions. This choice corresponds to the type II 2HDM, which is realized in the MSSM. The Higgs potential then takes the form \cite{HH}
\begin{equation}
\begin{aligned}
V(H_1,H_2)=&\left(m_{H_1}^2+|\mu|^2\right)|H_1|^2+\left(m_{H_2}^2+|\mu|^2\right)|H_2|^2-B\mu\left(\epsilon_{ij}H_1^iH_2^j+\rm{h.c.}\right)+\\
&\frac{1}{2}g^2\left|H_1^{i*}H_2^i\right|^2+\frac{1}{8}\left(g^2+g^{\prime 2}\right)\left(|H_1|^2-|H_2|^2\right)^2,
\end{aligned}
\label{eq:Higgspot}
\end{equation}
where $\epsilon_{ij}$ is the completely antisymmetric tensor, with $\epsilon_{12}=1$. The mass parameters $m_{H_1}^2$, $m_{H_2}^2$ may take on negative values in order to break the electroweak (EW) symmetry. A non-zero value for $B\mu$ breaks softly the discrete symmetry implemented for the type II model. The quartic terms in the MSSM Higgs potential are determined by the EW gauge couplings $g$ and $g^\prime$. 
In the MSSM, all parameters in the Higgs potential are real, and $\cp$ is conserved in the Higgs sector at tree level.

 Assigning vacuum expectation values to the neutral Higgs components we define, as usual, $\tan\beta\equiv\Ave{H_2^0}/\Ave{H_1^0}\equiv v_2/v_1$. Fixing $v^2=v_1^2+v_2^2$ to give the correct value for $m_Z$, the MSSM Higgs sector is completely determined at tree level by two parameters: $\tan\beta$ and one common mass scale for the Higgs bosons. The physical charged Higgs is defined through $H^\pm = -H_1^\pm \sin\beta +H_2^\pm\cos\beta$. Since there is only one free mass parameter, the masses of the neutral and charged Higgs bosons are related by tree level mass relations. For our purposes, the most important such relation is
\begin{equation}
 m^2_{H^+}=m^2_A+m^2_W,
\label{eq:massrel}
\end{equation}
between $m_{H^+}$ and the mass $m_A$ of the $\cp$-odd Higgs boson. This relation shows directly the quasi-degeneracy of $m_{H^+}$ and $m_A$ for large Higgs masses.

The charged Higgs phenomenology in the MSSM differs from that of the general 2HDM in a few respects: the presence of the EW gauge couplings in Eq.~(\ref{eq:Higgspot}) ensures that tree level unitarity is fulfilled. The additional theoretical constraints lead to mass relations such as Eq.~(\ref{eq:massrel}). There are also new aspects related to the additional states introduced by supersymmetry. Although the flavor structure in the MFV framework is left intact, sparticles may contribute indirectly to the same flavor observables as charged Higgs through loop effects. New decay chains where $H^+$ is produced, or new decay modes for $H^+$ into SUSY states, can also appear.

\subsection{SUSY Yukawa corrections}
At loop level, the discrete symmetry is broken in the Yukawa sector, inducing couplings to the ``wrong'' Higgs doublet \cite{Carena:1994bv,Hall&al:PRD:1994,Carena:NPB:2000}. Some of the induced corrections are enhanced by $\tan\beta$, and for precision phenomenology it is important to take them into account. Using an effective Lagrangian approach, the charged Higgs coupling to fermions is modified as follows \cite{Buras:2002vd}
\begin{equation}
V_{ij} m_{d_{j}}\tan\beta \to V_{ij}^{\rm{eff}}\frac{\overline{m}_{d_{j}}\tan\beta}{1+\tilde{\epsilon}_i\tan\beta},
\end{equation}
where $\overline{m}_{d_{j}}$ and $V_{ij}^{\rm{eff}}$ refers to experimental quantities.
Assuming perturbation theory remains valid, $|\tilde{\epsilon}_i\tan\beta|<1$. 
The correction $\tilde{\epsilon}_i$ is composed from the two pieces
\begin{equation}
\tilde{\epsilon}_i=\epsilon_0+\epsilon_2\delta_{i3},
\label{eq:epsilonb}
\end{equation}
where the second term, proportional to the top Yukawa coupling $y_t$, only contributes to the couplings involving the top quark, i.e. $tb$, $ts$, and $td$.
Explicitly, the $\tan\beta$ enhanced corrections are \cite{DAmbrosio:2002ex}
\begin{equation}
\epsilon_0=-\frac{2\alpha_s}{3\pi}\frac{\mu}{\mg}H_2\left(\frac{m_Q^2}{\mg^2},\frac{m_D^2}{\mg^2}\right)
\label{eq:eps0}
\end{equation}
\begin{equation}
\epsilon_2=-\frac{y_t^2}{16\pi^2}\frac{A_t}{\mu}H_2\left(\frac{m_Q^2}{|\mu|^2},\frac{m_U^2}{|\mu|^2}\right),
\end{equation}
where $m_Q^2,m_U^2,m_D^2$ are generation dependent soft SUSY breaking scalar masses, and $A_t$ the trilinear stop coupling. The function $H_2(x,y)$ is defined as
\begin{equation}
H_2(x,y)=\frac{x\ln x}{(1-x)(x-y)}+\frac{y\ln y}{(1-y)(y-x)}.
\end{equation}
For the third generation Yukawa correction, the compact notation
\begin{equation}
\Delta_b\equiv\tilde{\epsilon}_3\tan\beta
\end{equation}
is sometimes used. Considering the decoupling limit, when the SUSY masses are sent to a common high scale, the expression for $\tilde{\epsilon}_i$ becomes particularly simple. Neglecting the second part of Eq.~(\ref{eq:epsilonb}), one obtains $\tilde{\epsilon}_i=\epsilon_0=\mathrm{sign}(\mu)\times\alpha_s/(3\pi)$. This limit also gives the simple estimate $|\epsilon_0|\simeq 0.01$ for the magnitude of the corrections.

In a renormalization group improved treatment, $\overline{m}_b$ should be renormalized at $\mu_R=m_{H^+}$ to include QCD effects to all orders \cite{Braaten&Leveille:PRD:1980}, while $\tilde{\epsilon}_i$ is preferentially evaluated at the SUSY breaking scale $M_\mathrm{SUSY}$ \cite{Carena:NPB:2000} to avoid large logarithms of the type $\log ( \mu_R/M_\mathrm{SUSY})$.\footnote[1]{The recently completed two loop calculation indicates a change of $\mathcal{O}(10-15\%)$ in $\tilde{\epsilon}_3$ \cite{Noth:2008tw}. Both the sign and magnitude of the correction depends on the MSSM scenario. The renormalization scale dependence is significantly reduced compared to the one loop result.}

\section{Specification of the MSSM models}
\label{sect:susymodels}
We consider two specific scenarios to illustrate the constraints and collider prospects for charged Higgs bosons in the MSSM. Both models assume SUSY breaking mediated by gravity, minimal flavor violation (MFV), and conservation of $\mathcal{R}$--parity. The first model is the constrained MSSM (CMSSM), characterized by the set of parameters $\{m_0,m_{1/2},A_0,\tan\beta,\mathrm{sgn}(\mu)\}$. Here $m_0$ is the universal mass of the scalars, $m_{1/2}$ the universal gaugino mass, $A_0$ the universal trilinear coupling, and $\tan\beta$ the ratio of the vacuum expectation values of the Higgs doublets, as given above. The CMSSM model invokes unification boundary conditions at a very high scale $m_\mathrm{GUT}$ where the universal mass parameters are specified. The masses at the EW scale are determined through renormalization group evolution. Additionally, the radiative corrections must generate the correct shape of the Higgs potential in order to break the EW symmetry.

The second model we consider involves non-universal Higgs masses (NUHM). This model generalizes the CMSSM, allowing for the GUT scale mass parameters of the Higgs doublets to have values different from $m_0$, i.e. $m_{H_1}\neq m_{H_2}\neq m_0$. These two additional parameters with dimension of mass can be traded for two other parameters at a lower scale, conveniently the $\mu$ parameter and the mass $m_A$ of the $\cp$-odd Higgs boson. The tree level mass relation (\ref{eq:massrel}) then implies that the charged Higgs boson mass can be treated essentially as a free parameter, an important difference comparing to the CMSSM.

To investigate the parameter spaces of the CMSSM and NUHM, we perform scans in a random grid generating of order $10^5$ points for each scenario. The input ranges used for the parameters are given in Table~\ref{tab:NUHM}.
\begin{table}
\centering
\begin{tabular}{|c|ccl|}
   \hline                
         Parameter  & min & max & note\\
         \hline
         $m_0$  & $50$ & $2000$ &\\[3pt]
         $m_{1/2}$  & $50$ & $2000$ &\\[3pt]
         $A_0$  & $-2000$ & $2000$ &\\[3pt]
         $\mu$  & $-2000$ & $2000$ & CMSSM: only sign $\pm$\\[3pt]
         $m_A$  & $5$ & $600$ & NUHM only\\[3pt]
         $\tan\beta$  & $1$ & $60$ & \\[3pt]
         \hline
\end{tabular}
 \caption{Parameter ranges used for the CMSSM and NUHM scans. Dimensionful values (all parameters except $\tan\beta$) are given in GeV.}
\label{tab:NUHM}
\end{table}

The spectrum of SUSY particle masses and couplings is calculated for each model point using SOFTSUSY 2.0.18 \cite{Allanach:2001kg}. Full 2-loop RGE evolution is employed for all MSSM parameters, Yukawa and gauge couplings. The scale at which the universal MSSM parameters are specified is determined by unification of the EW gauge couplings.

In experimental simulation studies, as performed both by ATLAS and CMS, an updated version of the $m_h$--max scenario \cite{Carena:1999xa} is used as a common benchmark. This scenario is phenomenologically defined by the weak scale parameters
\[
\begin{aligned}
M_\mathrm{SUSY}&=1\, \mathrm{TeV} \\
M_2&=200\, \mathrm{GeV} \\
M_3&=800\, \mathrm{GeV} \\
X_t^{\mathrm{OS}}&=2\, \mathrm{TeV} \\
\mu&=200\, \mathrm{GeV}.
\end{aligned}
\]
The gaugino masses have values inspired by gauge coupling unification, and $M_1=5/3 M_2\tan^2\theta_W $. $X_t^{\rm{OS}}$ is defined in the on-shell scheme and relates the amount of mixing in the stop sector to the trilinear coupling $A_t$ through $X_t^{\rm{OS}}=A_t-\mu\cot\beta$. Compared to the original $m_h$--max scenario, the preferred sign of $\mu$ has later been changed to positive. As we will see below, this accommodates better for the experimental results on the anomalous magnetic moment of the muon.

\section{Constraints}
\label{sect:constraints}
We use a set of direct and indirect constraints in order to determine the parameter space regions allowed for charged Higgs bosons in the CMSSM and NUHM scenarios. Present data already provide interesting information on the models, while improvements in the theoretical calculations of both the Standard Model and supersymmetric contributions to a number of low energy observables increase their predictability. Some recent analyses showing constraints on the MSSM parameter space can be found in \cite{Carena:2006ai,Mahmoudi:2007gd,Domingo:2007dx,Ellis:2007fu,Ellis:2007ss,Heinemeyer:2008fb,Allanach:arxiv:2008}.

To obtain constraints on ($m_{H^+},\tan\beta$) we consider a) direct mass limits from LEP and the Tevatron, b) flavor data constraints, c) the muon anomalous magnetic moment, and d) the dark matter relic density.
The flavor observables and $(g-2)_\mu$ are calculated with SuperIso v2.3 \cite{Mahmoudi:2007vz,SuperIso}. A brief description for each observable is given below, and a more detailed account of the calculations can be found in \cite{SuperIso}. MicrOMEGAs 2.1 \cite{Belanger:2006is,Belanger:2008sj} is used for computing the dark matter relic density. For brevity, we illustrate the exclusion by different constraints using the NUHM model points only. The combined constraints are presented both for the CMSSM and the NUHM models.

\subsection{Direct mass limits}
The non-observation of the charged Higgs boson, or any other SUSY particle, in direct search experiments at LEP2 and the Tevatron sets stringent limits on the masses of these particles. In Table~\ref{tab:dirlim} we have compiled a list of mass constraints from the PDG \cite{Amsler:2008zz}. Some of these limits are subject to auxiliary conditions, e.g. $\tan\beta<40$, which we take into account consistently.
\begin{table}[h!]
\centering
\begin{tabular}{|cccccccccccc|}
   \hline                
         Particle & $H^+$ &  $h$ & $\chi^0_1$ & $\chi^+_1$& $\tilde{e}_R$ & $\tilde{\mu}_R$ & $\tilde{\tau}_1$ &$\tilde{\nu}$ & $\tilde{b}_1$ &$\tilde{t}_1$&$\tilde{g}$\\
   \hline                

        Mass limit (GeV) & $79.3$ & $111$ & $46$ & $94$& $73$ & $94$ & $81.9$ & $94$ & $89$ & $95.7$ & $308$\\
         \hline
\end{tabular}
 \caption{Lower limits at $95\%$ C.L.~on masses of sparticles and MSSM Higgs bosons. Some limits are subject to auxiliary conditions which are not listed in the table. We refer to \cite{Amsler:2008zz} for the complete list and further details on how they are obtained.}
\label{tab:dirlim}
\end{table}

The limit on the mass $m_h$ of the lightest $\cp$-even Higgs boson is very important for constraining also the heavier Higgs bosons, including $m_{H^+}$. This is a consequence of the tree level relations between the Higgs masses in the MSSM. In this work we apply the SM limit $m_h\gtrsim 114$~GeV \cite{Barate:2003sz}. Assigning a $3$ GeV intrinsic uncertainty in the Higgs mass prediction from higher order corrections \cite{Allanach:2004rh}, the value we finally use is $m_h>111$ GeV. As discussed in \cite{Schael:2006cr,Asano:2007gv,Ellis:2007fu}, it is possible in the MSSM to lower this bound down to $m_h\simeq 90$~GeV, in particular for high $\tan\beta$, by reducing the $ZZh$ coupling. The precise value for this limit depends on the MSSM scenario, requiring a complete reanalysis of the LEP Higgs search results for each model point, which is beyond the scope of the present work.
\begin{figure}[!t]
\begin{center}
\includegraphics[width=7.2cm,keepaspectratio]{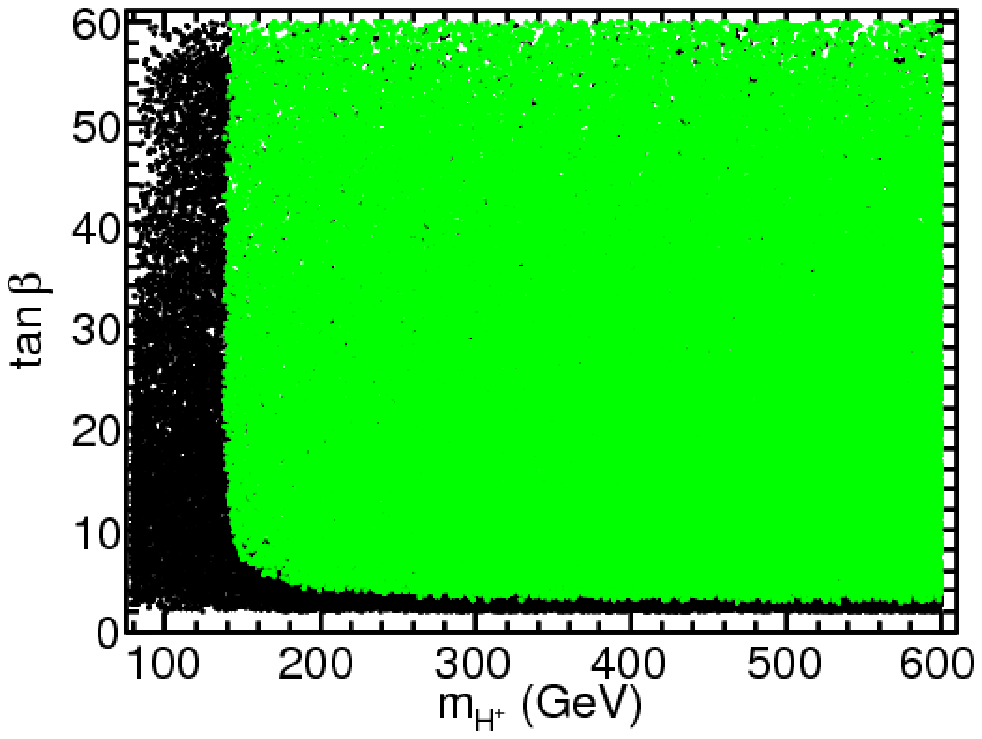}\hspace*{1cm}
\includegraphics[width=7.2cm,keepaspectratio]{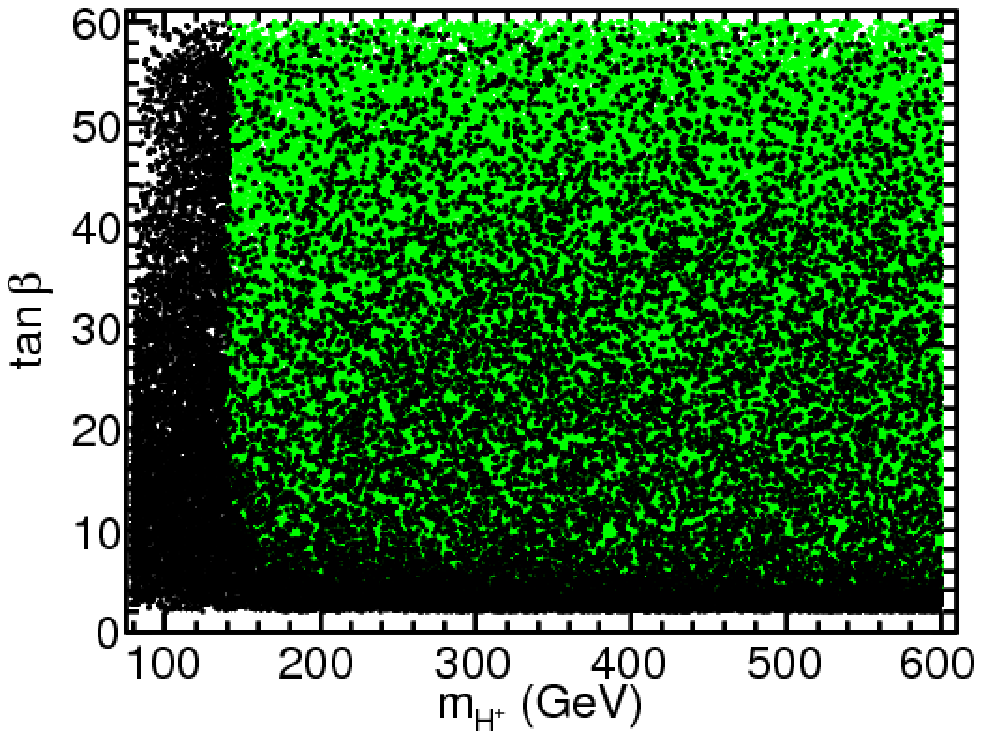}
\end{center}
\caption{Constraints from direct mass limits on the six dimensional NUHM parameter space, projected onto the plane $(m_{H^+},\tan\beta)$. The left plot shows the allowed points (green) in the foreground, whereas the right plot shows instead the excluded points (black) in the foreground. The difference illustrates the dependence on the other NUHM parameters.}
\label{fig:direct}
\end{figure}%

Figure~\ref{fig:direct} shows the effect in the $(m_{H^+},\tan\beta)$ plane of imposing the direct mass limits on the NUHM model points. The scan over the six dimensional parameter space is treated as follows: in the left plot, the allowed points are displayed in the foreground. The opposite is true for the right plot, where the excluded points are shown in the foreground. In this way the dependence on the model parameters which are not shown in the figure can be determined by comparison. A large difference between the two plots in a particular region indicates strong dependence on the other parameters, whereas a small or no difference indicates weak dependence.
 
Some distinct features of the direct mass constraints are visible in Figure~\ref{fig:direct}. There appears a lower limit of $m_{H^+}\gtrsim 135$ GeV, independent of $\tan\beta$. This value follows directly from the mass relation in Eq.~(\ref{eq:massrel}) with $m_A=111$~GeV. The higher order corrections to this relation are typically small in the NUHM models. This is not necessarily true in the more general MSSM, where special parameter regions with light squarks ($|\mu|>4M_{SUSY}$) allow for large mass splittings in the Higgs sector, even to the degree $m_{H^+}<m_A-m_W$  \cite{Eriksson:2006yt}.
 Lowering the limit of $m_h$, as aforementioned, would result in a corresponding shift in the $m_{H^+}$ limit. The region with $\tan\beta\lesssim 3$ is excluded, again as a result of the $m_h$ limit. Except in these two regions, the direct mass limits do not provide further constraints on $(m_{H^+},\tan\beta)$ in the NUHM models.

In the CMSSM, the direct mass limits are not more constraining than in the NUHM models. On the contrary, there is a region for high $m_{H^+}$ and $\tan\beta$ which is always allowed. This is because in the CMSSM at high $\tan\beta$, $m_0$ and $m_{1/2}$ cannot be too low, and  $h$, $\chi^0_1$ and $\tilde g$ are always sufficiently heavy to avoid the direct constraints.

\subsection{Flavor physics constraints}
Constraints on charged Higgs bosons can be obtained from low energy flavor physics experiments by measuring the decay rates of $B$ and $K$ mesons and comparing to the SM predictions. This is challenging in many respects: experimentally, because many of the interesting transitions are rare, and theoretically, since the predictions often suffer from large hadronic uncertainties. In the following, theoretical and experimental uncertainties are added in quadrature. The value $m_t=172.4$ GeV \cite{Group:2008vn} is used throughout.

\subsubsection{$b\to s\gamma$}
The rare FCNC process $b\to s\gamma$, occurring first at one loop level in the SM, allows for new physics contributions from a charged Higgs loop to be of comparable magnitude. Since the charged Higgs always contributes positively to this branching ratio, it is an effective tool to probe the 2HDM.
In the MSSM however, there exist additional contributions from loops with charginos and squarks, which may be either negative or positive. Hence the charged Higgs constraints in the MSSM are not necessarily as strict as those obtained in the pure 2HDM.

Following \cite{Misiak:2006ab}, the theoretical prediction for the inclusive branching ratio of $b\to s\gamma$ can be written
\begin{equation}
\rm{BR}(\bar{B} \to X_s \gamma)= \mathrm{BR}(\bar{B} \to X_c e \bar{\nu})_\mathrm{exp} 
\left| \frac{ V^*_{ts} V_{tb}}{V_{cb}} \right|^2 \frac{6 \alpha_{\rm em}}{\pi C} \left[ P(E_0) + N(E_0) \right],
\end{equation}
where $P(E_0)$ and $N(E_0)$ denote, respectively, the perturbative and non-perturbative contributions evaluated for a cut $E>E_0$ on the photon energy. $C$ is a semi-leptonic normalization factor. The details are described in \cite{Misiak:2006ab}. The SM prediction for this decay is known to NNLO accuracy \cite{Misiak:2006ab,Becher:2006pu,Gambino:2008fj}. Using the updated input values of \cite{Amsler:2008zz}, we obtain
\begin{equation}
\rm{BR}(\bar{B} \to X_s \gamma)_{\rm{SM}} = (3.06\pm0.22)\times10^{-4},
\end{equation}
while one would retrieve $\rm{BR}(\bar{B} \to X_s \gamma)_{\rm{SM}}=3.15\times10^{-4}$ with the input values of \cite{Misiak:2006ab}.
The combined experimental value of the branching ratio is updated by the Heavy Flavor Averaging Group (HFAG) \cite{Barberio:2008fa}. Their latest result is
\begin{equation}
\rm{BR}(\bar{B} \to X_s \gamma)_{\mathrm{exp}}=(3.52\pm0.23\pm0.09)\times10^{-4}.
\end{equation}
To obtain the allowed range for $\rm{BR}(\bar{B} \to X_s \gamma)$ in the MSSM, we follow the procedure outlined in \cite{Mahmoudi:2007gd,Ellis:2007fu}, in which the intrinsic MSSM uncertainty is added to the uncertainties of the experimental value and the SM prediction. The resulting allowed range at 95\% C.L. is
\begin{equation}
2.15\times 10^{-4}\leq \rm{BR}(\bar{B} \to X_s \gamma) \leq 4.89\times 10^{-4}.
\end{equation}

Another constraining observable which can be extracted from $b\to s\gamma$ transitions is the degree of isospin asymmetry. It has been shown earlier \cite{Mahmoudi:2007gd,Ahmady:2006yr}, that it often provides stricter limits on the parameters of different MSSM scenarios than the inclusive branching ratio. 
The isospin asymmetry $\Delta_0$ in the exclusive decay $B \to K^* \gamma$ is defined as
\begin{equation}
\Delta_{0\pm}\equiv\dfrac{\Gamma(\bar B^0\to\bar K^{*0}\gamma) - \Gamma(B^\pm \to K^{*\pm}\gamma)}{\Gamma(\bar B^0\to\bar K^{*0}\gamma) + \Gamma(B^\pm\to K^{*\pm}\gamma)}.
\end{equation} 
It can be determined from \cite{Kagan:2001zk}
\begin{equation}
\Delta_{0} = \mathrm{Re}(b_d-b_u),
\end{equation} 
where 
\begin{equation}
b_q = \dfrac{12\pi^2 f_B\,Q_q}{m_b\,T_1^{B\to K^*} a_7^c} \left( \dfrac{f_{K^*}^\perp}{m_b}\,K_1
   + \dfrac{f_{K^*} m_{K^*}}{6\lambda_B m_B}\,K_2 \right)
\end{equation}
are spectator quark dependent terms. We refer to \cite{Kagan:2001zk,Ahmady:2006yr} for the definition of the different terms, and their expressions in terms of Wilson coefficients.

Combining the most recent experimental values of BaBar \cite{Aubert:2008cy} and the results of Belle \cite{Nakao:2004th}, including the experimental and theoretical uncertainties, the allowed range
\begin{equation}
-1.7\times 10^{-2} < \Delta_0 < 8.9\times 10^{-2}
\end{equation}
is obtained at 95\% C.L. \cite{SuperIso}.
\begin{figure}[!t]
\begin{center}
\includegraphics[width=7cm,keepaspectratio]{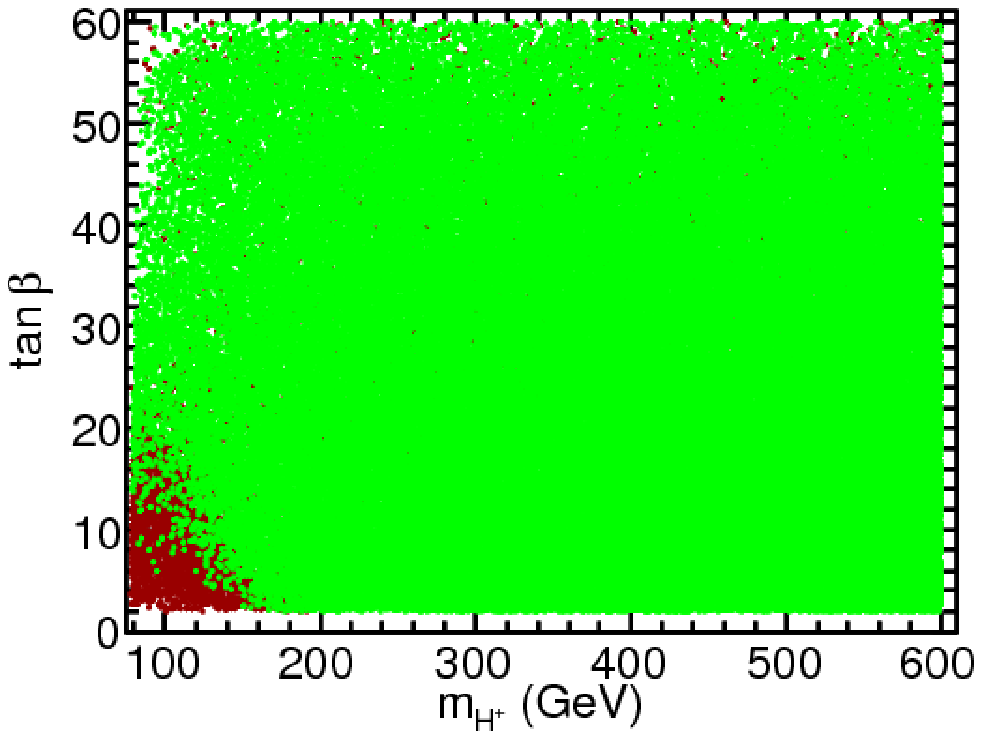}\hspace*{1cm}
\includegraphics[width=7cm,keepaspectratio]{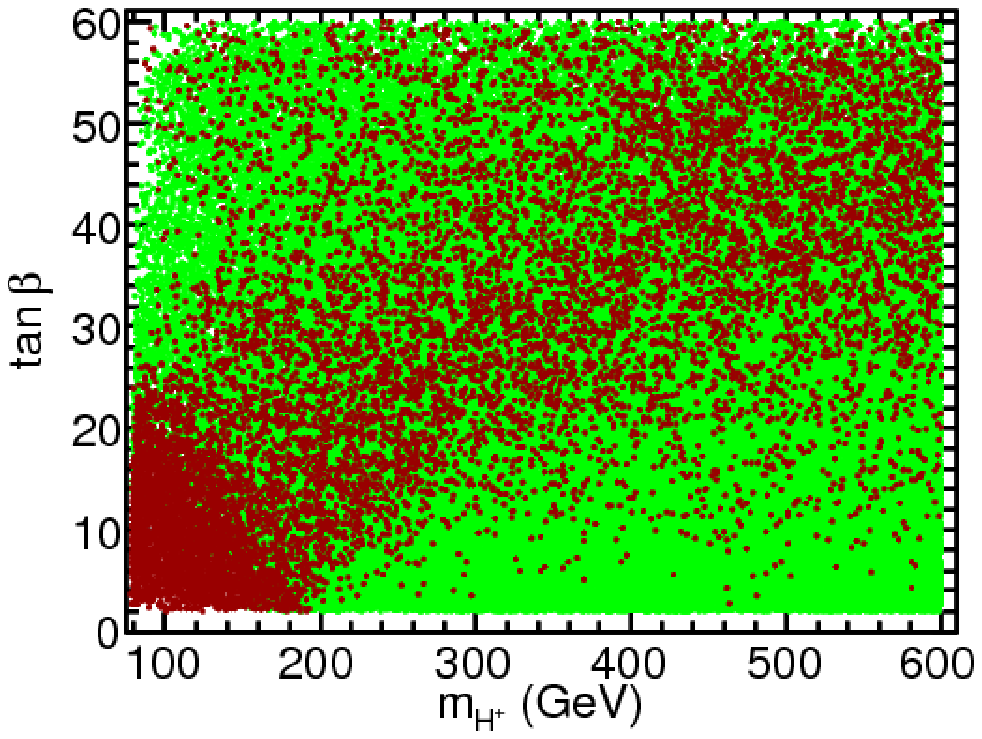}
\end{center}%
\caption{Constraints from the branching ratio and isospin asymmetry in $b\to s\gamma$ transitions on the six dimensional NUHM parameter space, projected onto the plane $(m_{H^+},\tan\beta)$. The left plot shows the allowed points (green) in the foreground, whereas the right plot shows instead the excluded points (red) in the foreground.}
\label{fig:bsg}
\end{figure}%

In the same way as for the direct constraints, the results of applying the $b\to s\gamma$ constraints to the NUHM model points are displayed in two separate plots in Figure~\ref{fig:bsg}. The figure includes both the constraints from the branching ratio and the isospin asymmetry combined. Low values for $m_{H^+}$ and $\tan\beta$ simultaneously are excluded, regardless of the other NUHM parameters. As expected however, it is possible in most of the parameter space to balance out the contributions from charged Higgs and from chargino--squarks against each other to a sufficiently high degree to be consistent with the SM. Observables based on $b\to s\gamma$ transitions thus represent high sensitivity to the MSSM parameters.

We find that it is easier in NUHM models to avoid the constraints from $b\to s\gamma$ than it is in the CMSSM. The isospin asymmetry is more restrictive in the CMSSM than the constraint from the branching ratio, and the main allowed region in the CMSSM is obtained for high $\tan\beta$.

\subsubsection{$B_u\to\tau\nu_\tau$}
In contrast to the $b \to s\gamma$ transitions, where the charged Higgs participates in loop diagrams, the process $B_u\to\tau\nu_\tau$ can be mediated by $H^+$ already at tree level. Since this decay is helicity suppressed in the SM, whereas there is no such suppression for the scalar $H^+$ exchange in the limit of high $\tan\beta$, these two contributions can be of similar magnitude \cite{Hou:1992sy}.

The leading order SM prediction for this decay is
\begin{equation}
\mathrm{BR}(B_u\to\tau\nu_\tau)_\mathrm{SM}=\frac{G_F^2f_B^2|V_{ub}|^2}{8\pi\Gamma_B}m_Bm_\tau^2\left(1-\frac{m_\tau^2}{m_B^2}\right)^2,
\label{eq:Btaunu}
\end{equation}
while the new physics contribution from $H^+$ is expressed through the ratio \cite{Akeroyd:2003zr}
\begin{equation}
\label{eq:RMSSM}
R^{\mathrm{MSSM}}_{\tau\nu_\tau}\equiv\frac{\mathrm{BR}(B_u\to\tau\nu_\tau)_{\mathrm{MSSM}} }{\mathrm{BR}(B_u\to\tau\nu_\tau)_{\mathrm{SM}}}=\left[1-\left(\frac{m_B^2}{m_{H^+}^2}\right)\frac{\tan^2\beta}{1+\epsilon_0\tan\beta}\right]^2.
\end{equation}
The leading SUSY-QCD corrections are included in this expression through $\epsilon_0$. Using $f_B=200\pm20$ MeV \cite{Lubicz:2008am}, and the combined value $|V_{ub}|=(3.95\pm0.35)\times10^{-3}$ \cite{Amsler:2008zz}, the SM branching ratio evaluates numerically to
\begin{equation}
\mathrm{BR}(B_u\to\tau\nu_\tau)_\mathrm{SM}=(1.10\pm 0.29)\times 10^{-4}.
\end{equation}
The SM prediction is compared to the current HFAG value \cite{Barberio:2008fa}
\begin{equation}
\mathrm{BR}(B_u \to \tau\nu_\tau)_\mathrm{exp}=(1.41\pm 0.43)\times 10^{-4}
\end{equation}
by forming the ratio
\begin{equation}
R_{\tau\nu_\tau}^{\mathrm{exp}}\equiv \frac{\mathrm{BR}(B_u\to\tau\nu_\tau)_{\mathrm{exp}} }{\mathrm{BR}(B_u\to\tau\nu_\tau)_{\mathrm{SM}}} =1.28\pm 0.38.
\label{Rtaunu} 
\end{equation}
\begin{figure}[!t]
\begin{center}
\includegraphics[width=7cm,keepaspectratio]{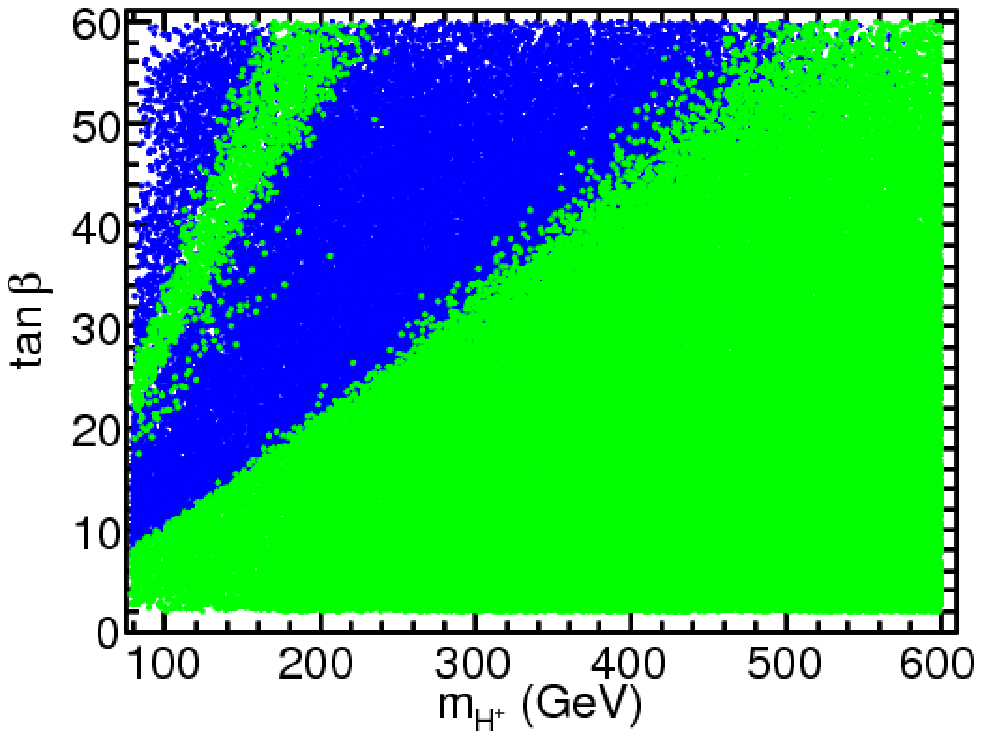}\hspace*{1cm}
\includegraphics[width=7cm,keepaspectratio]{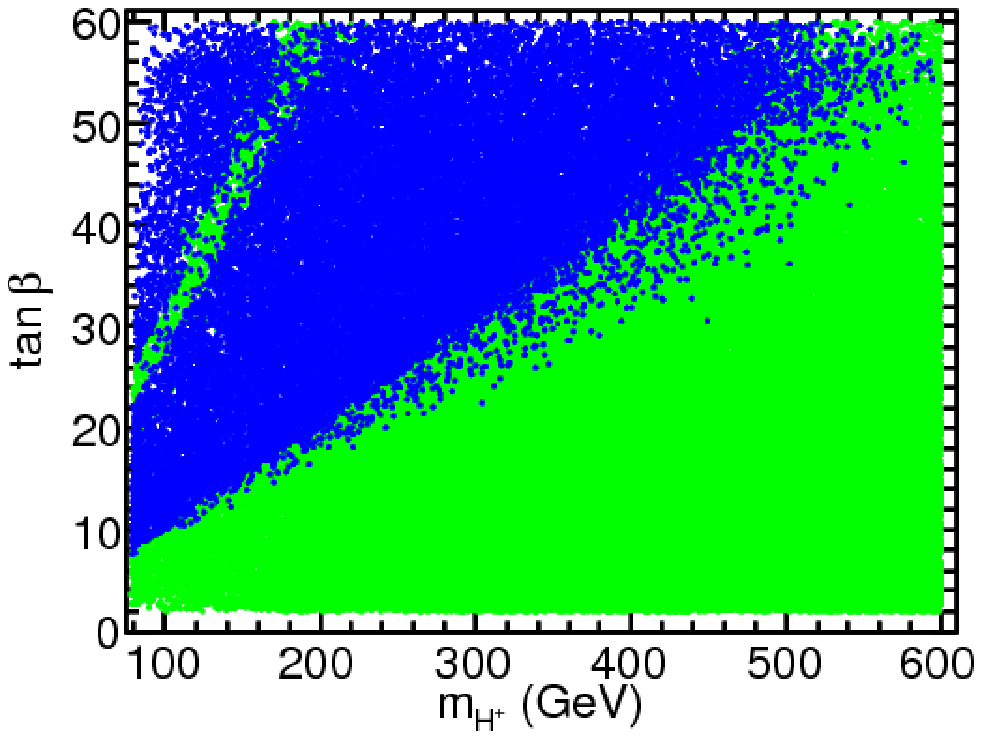}
\end{center}
\caption{Constraints from $\BR{B}{\tau\nu_\tau}$ on the six dimensional NUHM parameter space, projected onto the plane $(m_{H^+},\tan\beta)$. The left plot shows the allowed points (green) in the foreground, whereas the right plot shows instead the excluded points (blue) in the foreground.}
\label{fig:Btaunu}
\end{figure}%
This ratio still suffers large uncertainties from the determination of $|V_{ub}|$, since different measurements of this quantity are incompatible. The constraints obtained should therefore be treated merely as an indication, rather than as a strict limit on the same level as, for example, $b\to s\gamma$ transitions. Requiring $R^{\mathrm{MSSM}}_{\tau\nu_\tau}$ to be within $2\,\sigma$ of $R^{\mathrm{exp}}_{\tau\nu_\tau}$ provides the following allowed range:
\begin{equation}
0.53 < R^{\rm{MSSM}}_{\tau\nu_\tau} < 2.03
\end{equation}
where we have estimated the residual MSSM errors to be negligible.

The resulting constraints from $B_u\to \tau \nu_\tau$ are illustrated in Figure~\ref{fig:Btaunu}. It can be seen that a fairly large share of the available parameter space is affected. The allowed points fall in two disjoint regions. At low $m_{H^+}$ and large $\tan\beta$, the $H^+$ contributes twice the SM amplitude with opposite sign [see Eq.~(\ref{eq:RMSSM})]. The exclusion power varies only weakly between different MSSM models, as shown by the similarity of the left and right plots in Figure~\ref{fig:Btaunu}. This results from the fact that the only source of significant MSSM scenario dependence in this tree level observable is through the $\epsilon_0$ corrections. That this is a tree level observable also means that the results can be carried over essentially unchanged to the CMSSM, or to any MSSM model with MFV and $\mathcal{R}$-parity conservation.

\begin{figure}[!t]
\begin{center}
\includegraphics[width=4.8cm,keepaspectratio]{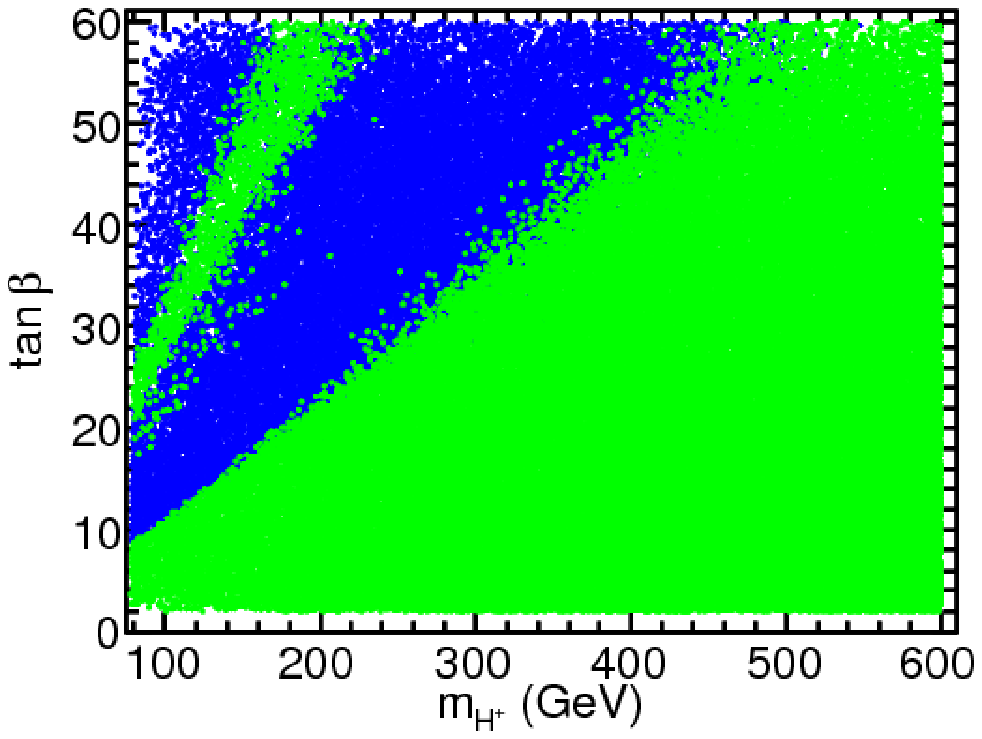}
\includegraphics[width=4.8cm,keepaspectratio]{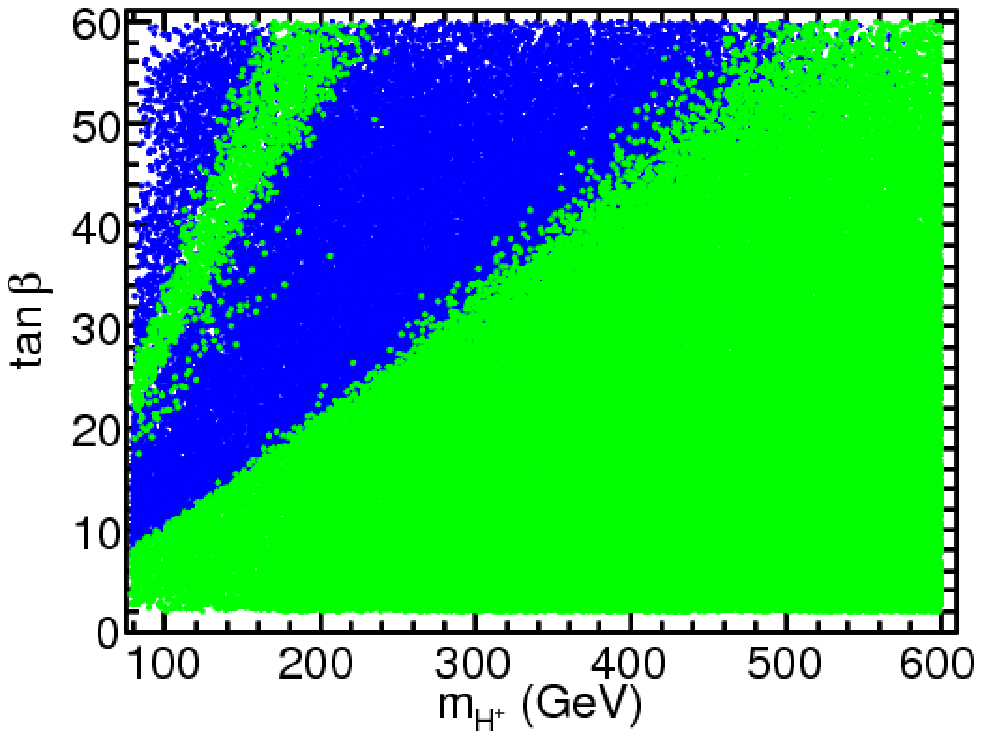}
\includegraphics[width=4.8cm,keepaspectratio]{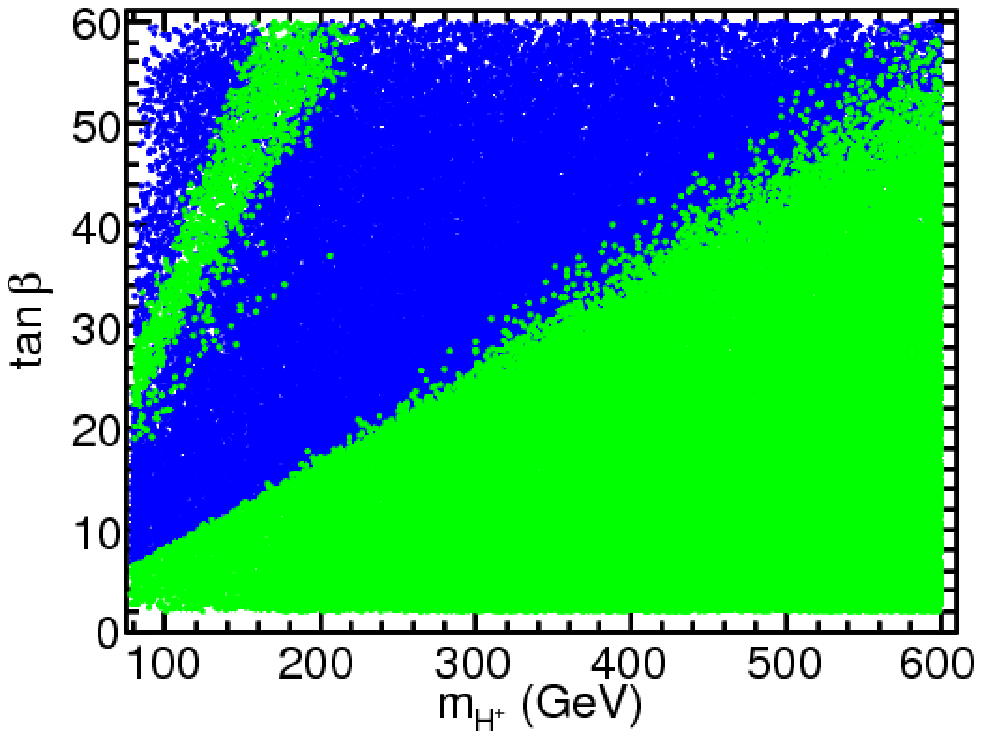}
\end{center}
\caption{Constraints on NUHM parameters obtained from $\BR{B_u}{\tau\nu_\tau}$ for three values of $|V_{ub}|$ from \cite{Amsler:2008zz}. Allowed points shown in green and excluded points in blue. From left to right $|V_{ub}^{\rm{incl}}|=(4.12\pm0.43)\times 10^{-3}$, $|V_{ub}^{\rm{comb}}|=(3.95\pm 0.35)\times 10^{-3}$, and $|V_{ub}^{\rm{excl}}|=(3.5\pm0.6)\times 10^{-3}$.}
\label{fig:comp_Btaunu}
\end{figure}

Treating the difference in $|V_{ub}|$ determination as a theoretical uncertainty would eliminate the constraint from $\BR{B_u}{\tau\nu_\tau}$. As an alternative, we present in Figure~\ref{fig:comp_Btaunu} the resulting constraints on $(m_{H^+},\tan\beta)$ obtained using three different values for $|V_{ub}|$,  corresponding to two separate determinations and to the combined value used for Figure~\ref{fig:Btaunu} above.

\subsubsection{$B\to D\tau\nu_\tau$}
Compared to $B_u\to\tau\nu_\tau$, the semi-leptonic decays $B \to D\ell\nu$  \cite{Grzadkowski:1991kb,Nierste:2008qe,Kamenik:2008tj} have the advantage of depending on $|V_{cb}|$, which is known to greater precision than $|V_{ub}|$. In addition, the $\mathrm{BR}(B \to D \tau \nu_\tau)$ is about $50$ times larger than $\mathrm{BR}(B _u\to \tau \nu_\tau)$ in the SM. The experimental determination remains however very complex due to the presence of at least two neutrinos in the final state. The branching ratio, including the SM and charged Higgs contributions, can be obtained from \cite{Kamenik:2008tj}
\begin{equation}
\begin{aligned}
\frac{d\Gamma(B\to D\ell \nu)}{dw} =&\, \frac{G_F^2|V_{cb}|^2 m_B^5}{192\pi^3}\rho_V(w) \\
& \times\left[1 - \frac{m_{\ell}^2}{m_B^2}\, \left\vert 1- t(w)\, \frac{m_b}{(m_b-m_c)m^2_{H^{+}}}\,\frac{\tan^2\beta}{1+\epsilon_0\tan\beta} \right\vert^2 \rho_S(w) \right], 
\label{bdtaunu}
\end{aligned}
\end{equation}
where the kinematic variable $w=v_D\cdot v_B$ is written in terms of the meson four velocities, and $t(w) = m_B^2 + m_D^2 - 2 w \,m_D \,m_B$.
The definitions of the scalar and vector form factors ($\rho_S$ and $\rho_V$, respectively) can be found in \cite{Kamenik:2008tj}. To reduce some of the theoretical uncertainties, the ratio
\begin{equation}
\xi_{D\ell\nu} \equiv \frac{\BR{B}{D\tau\nu_\tau}}{\BR{B}{D e \nu_e}}
\end{equation}
is considered, which is expected to be sensitive to charged Higgs contributions through the final state with a $\tau$ lepton. The SM prediction for this ratio is
\begin{equation}
\xi^{\rm{SM}}_{D\ell\nu}=(29\pm 3)\times 10^{-2},
\end{equation}
where the main uncertainty comes from the form factors \cite{Kamenik:2008tj}. The most recent experimental result by the BaBar collaboration is \cite{Aubert:2007dsa} 
\begin{equation}
\xi_{D\ell\nu}^{\mathrm{exp}} = (41.6 \pm 11.7 \pm 5.2) \times 10^{-2}.
\end{equation}
To derive the allowed range for this observable in the MSSM, the theoretical and experimental results are combined. Including also the enhancement of the form factor uncertainties by the presence of the charged Higgs contribution in Eq.~(\ref{bdtaunu}), we use the following interval at $95\%$ C.L. in our analysis
\begin{equation}
15.1 \times 10^{-2} < \xi_{D\ell\nu} < 68.1 \times 10^{-2}.
\end{equation}

\begin{figure}[!t]
\begin{center}
\includegraphics[width=7cm,keepaspectratio]{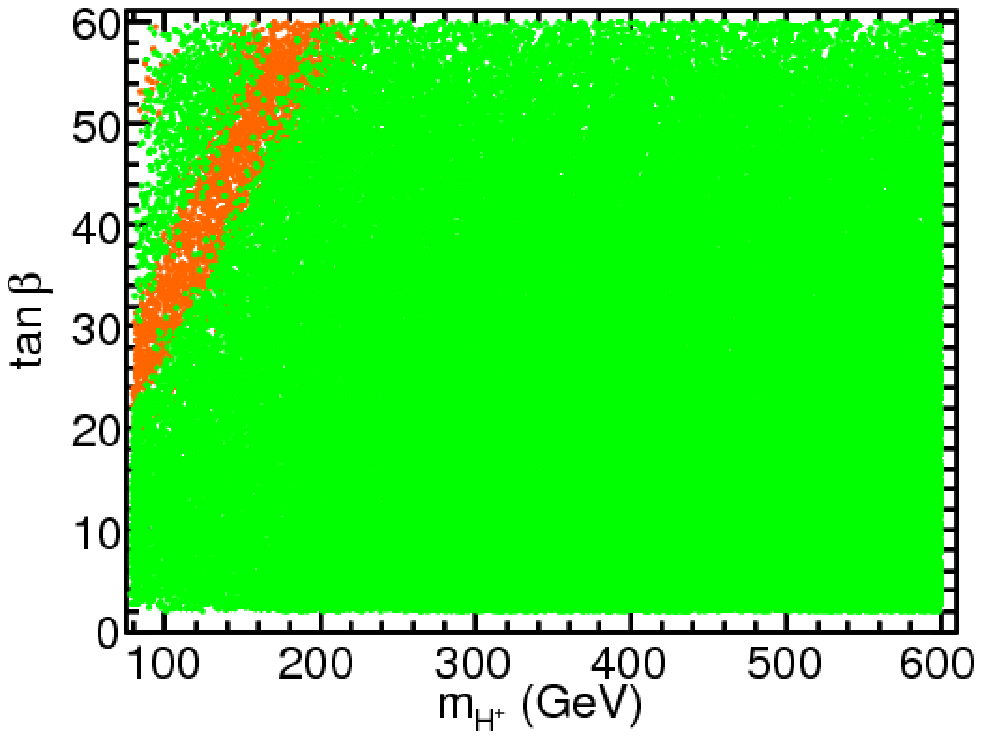}\hspace*{1cm}
\includegraphics[width=7cm,keepaspectratio]{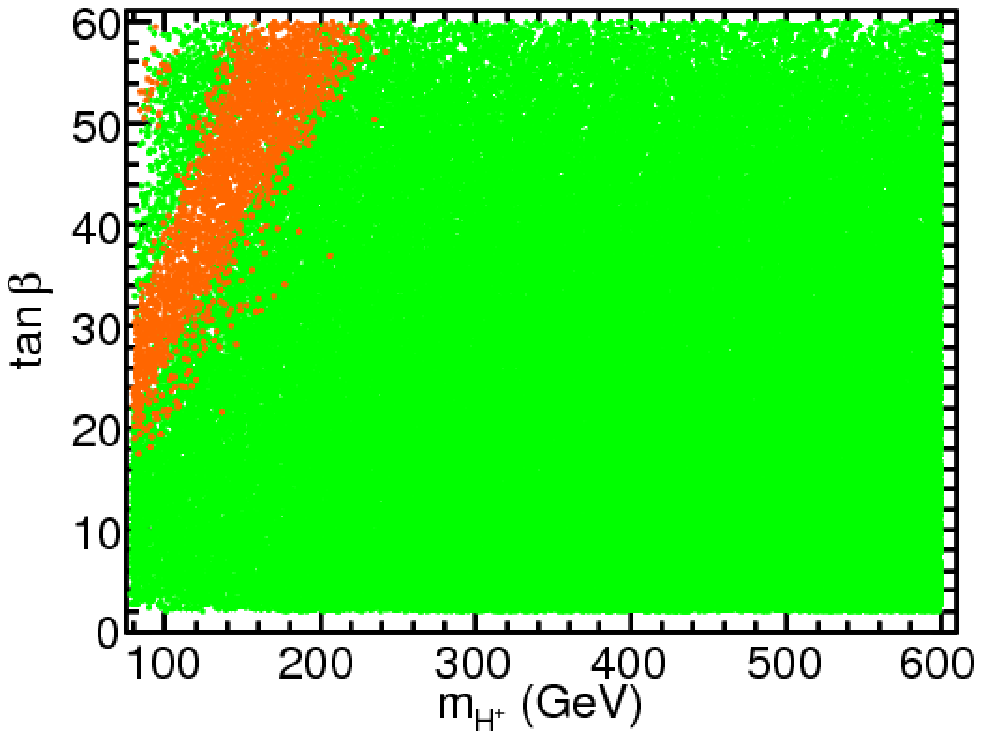}
\end{center}
\caption{Constraints from $\BR{B}{D\tau\nu_\tau}$ on the six dimensional NUHM parameter space, projected onto the plane $(m_{H^+},\tan\beta)$. The left plot shows the allowed points (green) in the foreground, whereas the right plot shows instead the excluded points (orange) in the foreground.}
\label{fig:BDtau}
\end{figure}

The results of applying the $B\to D\tau\nu_\tau$ constraint to the NUHM points are shown in Figure~\ref{fig:BDtau}. A narrow strip at large $\tan\beta$ is excluded for $m_{H^+}<200$ GeV. This excluded region provides excellent complementarity to the $B_u\to\tau\nu_\tau$ constraints obtained above, as most of the parameter space region still allowed at low $m_{H^+}$ is covered. Similar (small) MSSM model dependence through $\epsilon_0$ applies in this case as for $B_u\to \tau\nu_\tau$, and the result is unchanged in the CMSSM.

\subsubsection{$B_s\to\mu^+\mu^-$}
As a final $B$ meson decay we investigate the rare process $B_s \to \mu^+ \mu^-$, which has so far not been observed experimentally. At high $\tan\beta$, the MSSM contribution to this process is dominated by the exchange of neutral Higgs bosons. We therefore expect indirect constraints on $m_{H^+}$ and $\tan\beta$ from the MSSM mass relations. 
The $\BR{B_s}{\mu^+\mu^-}$ can be expressed as \cite{Bobeth:2001sq}
\begin{eqnarray}
\mathrm{BR}(B_s \to \mu^+ \mu^-) &=& \frac{G_F^2 \alpha^2}{64 \pi^3} f_{B_s}^2 \tau_{B_s} M_{B_s}^3 |V_{tb}V_{ts}^*|^2 \sqrt{1-\frac{4 m_\mu^2}{M_{B_s}^2}} \nonumber \\
&\times& \left\{\left(1-\frac{4 m_\mu^2}{M_{B_s}^2}\right) M_{B_s}^2 | C_S |^2 + \left |C_P M_{B_s} -2 \, C_A \frac{m_\mu}{M_{B_s}} \right |^2\right\} ,
\end{eqnarray}
where the coefficients $C_S$, $C_P$, and $C_A$ parametrize different contributions. Within the SM, $C_S$ and $C_P$ are small, whereas the main contribution entering through $C_A$ is helicity suppressed. In the MSSM, both $C_S$ and $C_P$ can receive large contributions from scalar exchange. The $B_s$ decay constant $f_{B_s}=245\pm 25$ MeV \cite{Lubicz:2008am} constitutes the main source of uncertainty in this expression. The SM prediction is
\begin{equation}
\mathrm{BR}(B_s \to \mu^+ \mu^-)_\mathrm{SM}=(3.2\pm 0.5) \times 10^{-9},
\end{equation}
while the current experimental limit, derived by the CDF collaboration, is \cite{Aaltonen:2007kv}:
\begin{equation}
\mathrm{BR}(B_s \to \mu^+ \mu^-) < 5.8 \times 10^{-8}
\end{equation}
at 95\% C.L. The experimental limit is thus still an order of magnitude away from the SM prediction, allowing for substantial SUSY contributions. Including theoretical uncertainties, we compare the MSSM prediction to the upper limit at 95\% C.L.
\begin{equation}
\mathrm{BR}(B_s \to \mu^+ \mu^-) < 6.6 \times 10^{-8}.
\end{equation}

\begin{figure}[!t]
\begin{center}
\includegraphics[width=7cm,keepaspectratio]{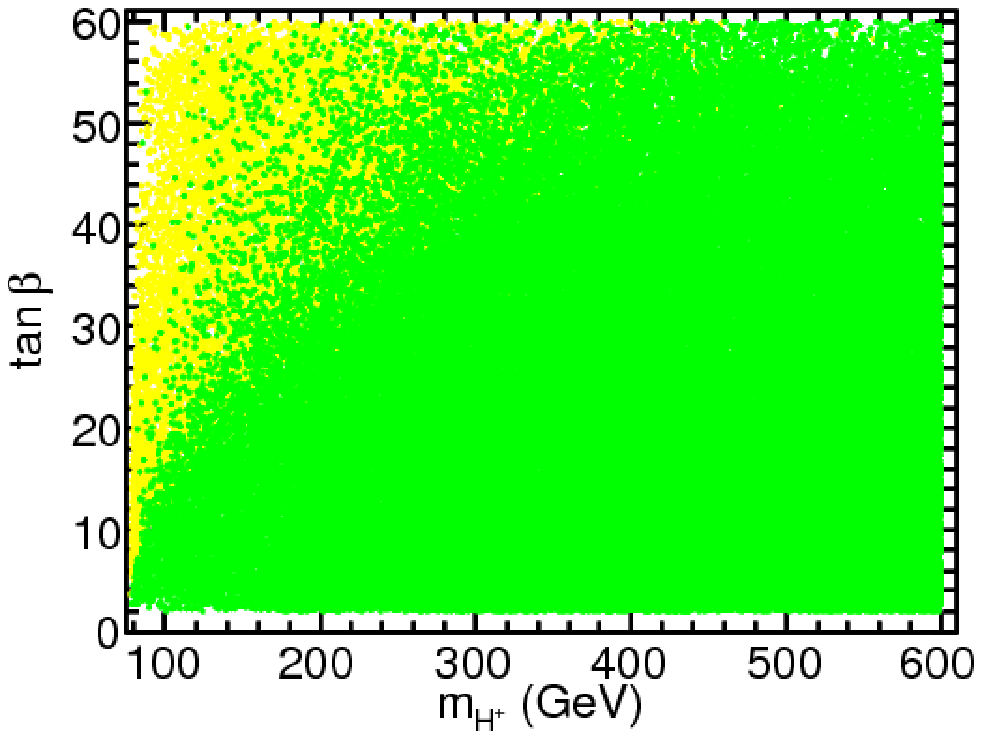}\hspace*{1cm}
\includegraphics[width=7cm,keepaspectratio]{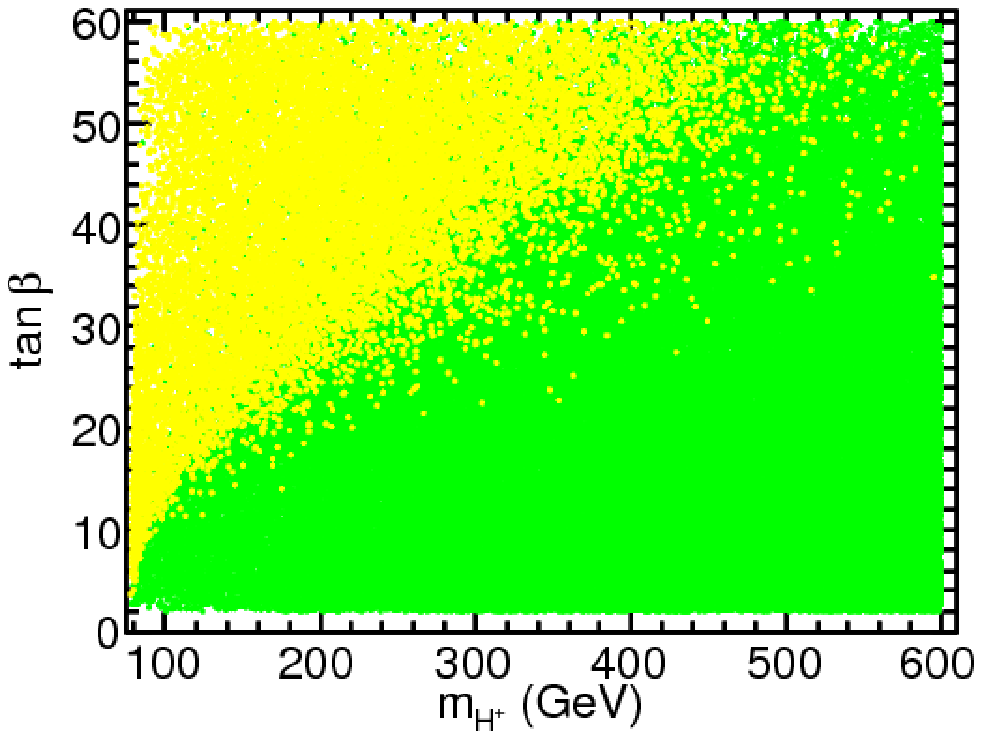}\\
\end{center}
\caption{Constraints from $\BR{B_s}{\mu^+\mu^-}$ on the six dimensional NUHM parameter space, projected onto the plane $(m_{H^+},\tan\beta)$. The left plot shows the allowed points (green) in the foreground, whereas the right plot shows instead the excluded points (yellow) in the foreground.}
\label{fig:Bmumu}
\end{figure}

The resulting constraints from $B_s\to\mu^+\mu^-$ are shown in Figure~\ref{fig:Bmumu}. The indirect dependence of this constraint on $m_{H^+}$ is clearly visible; in the right plot the possible exclusion is seen to be quite effective, whereas there is a large transition to allowed points going to the left plot. Hence the dependence on the MSSM scenario is large, and the constraints on ($m_{H^+},\tan\beta$) become dependent on the masses of the sparticles, for example the charginos. The only region which is almost completely excluded is for very small $m_{H^+}$ and large $\tan\beta$.

In the CMSSM, $B_s\to \mu^+ \mu^-$ removes most points with $\tan\beta>50$, complementing the direct constraints and $b\to s\gamma$.

\subsubsection{$K\to\mu\nu_\mu$}
Another decay process, which has many similarities to $B_u \to \tau \nu_\tau$, is $K \to \mu \nu_\mu$. This decay can also be mediated by the charged Higgs at tree level, although in this case the $H^+$ contribution is reduced by the coupling of $H^+$ to lighter quarks.
In order to reduce the theoretical uncertainties from $f_K$, the ratio of partial widths
\begin{eqnarray}
\dfrac{\Gamma(K \rightarrow \mu \nu_\mu)}{\Gamma(\pi \rightarrow \mu \nu_\mu)}&=& 
\left|\frac{V_{us}}{V_{ud}} \right|^2\frac{f^2_K m_K}
{f^2_\pi m_\pi}\left(\frac{1-m^2_\ell/m_K^2}{1-m^2_\ell/m_\pi^2}\right)^2 \nonumber \\
&& \times \left(1-\frac{m^2_{K^+}}{M^2_{H^+}}\left(1 - \frac{m_d}{m_s}\right)\frac{\tan^2\beta}{1+\epsilon_0\tan\beta}\right)^2 \left(1+\delta_{\rm em}\right)
\end{eqnarray}
is usually considered. Here $\delta_{\rm em} = 0.0070 \pm 0.0035$ is a long distance electromagnetic correction factor. As suggested in \cite{Antonelli:2008jg}, we study instead the quantity 
\begin{equation}
R_{\ell 23}\equiv\left| \frac{V_{us}(K_{\ell 2})}{V_{us}(K_{\ell 3})} \times \frac{V_{us}(0^+ \to 0^+)}{V_{ud}(\pi_{\ell 2})} \right|.
\end{equation}
Here $V_{us}(K_{\ell i})$ refers to $V_{us}$ as measured in leptonic decay of $K$ with $i$ particles in the final state (two leptons and a number of pions), and similarly for $V_{ud}$. The $0^+\to 0^+$ denotes nuclear beta decay.
In the SM $R_{\ell 23}=1$, while the contribution from charged Higgs in the MSSM attains the simple form
\begin{equation}
R_{\ell 23}=\left|1-\frac{m^2_{K^+}}{M^2_{H^+}}\left(1 - \frac{m_d}{m_s}\right)\frac{\tan^2\beta}{1+\epsilon_0\tan\beta}\right|.
\end{equation}
Using $m_d/m_s=1/20$ \cite{Amsler:2008zz}, the MSSM prediction can be directly compared to the experimental value \cite{Antonelli:2008jg}
\begin{equation}
R_{\ell 23}=1.004\pm 0.007.
\label{Rl23}
\end{equation}
In the extraction of this value, the ratio $f_K/f_\pi$ has been fixed to the value $f_K/f_\pi=1.189\pm 0.007$ obtained from lattice QCD using staggered quarks \cite{Follana:2007uv}. It should be noted that the uncertainty thus obtained for $R_{\ell 23}$ is most probably overly optimistic. Indeed, many approaches exist to determine $f_K/f_\pi$, and some reservation remains about staggered fermions  \cite{Creutz:2007rk}. If, for example, the value $f_K/f_\pi = 1.205\pm 0.018$ (obtained using the domain wall formulation \cite{Allton:2007hx}) is used instead, the $\BR{K}{\mu\nu}$ provides no constraints on the studied NUHM parameters. We therefore stress that the constraints obtained from (\ref{Rl23}) should serve only as an indication. 

\begin{figure}[!t]
\begin{center}
\includegraphics[width=7cm,keepaspectratio]{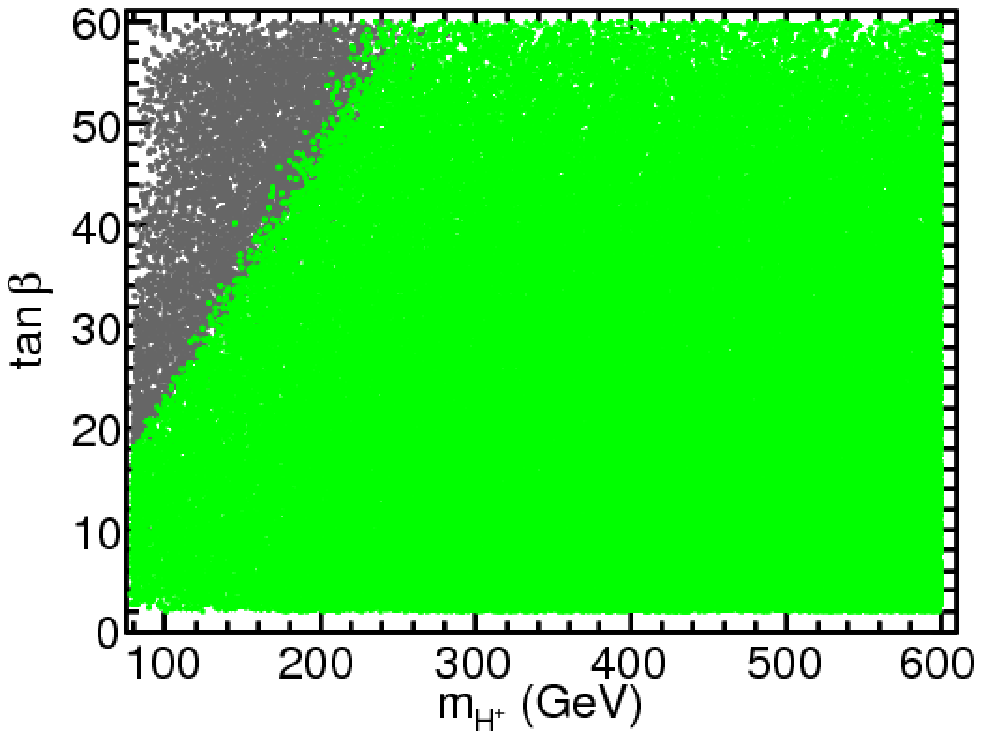}\hspace*{1cm}
\includegraphics[width=7cm,keepaspectratio]{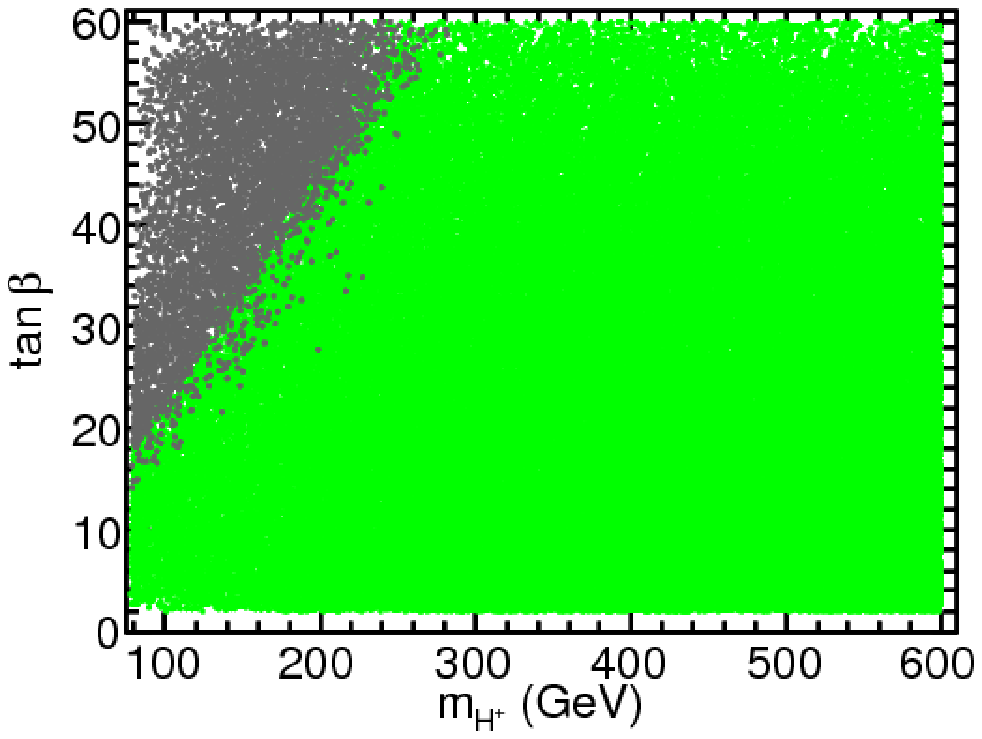}
\end{center}
\caption{Constraints from $K\to\mu\nu_\mu$ transitions on the six dimensional NUHM parameter space, projected onto the plane $(m_{H^+},\tan\beta)$. The left plot shows the allowed points (green) in the foreground, whereas the right plot shows instead the excluded points (gray) in the foreground. The value $f_K/f_\pi=1.189\pm 0.007$ was used in obtaining these constraints.}
\label{fig:Kmunu}
\end{figure}
As can be seen from Figure~\ref{fig:Kmunu}, more precise estimates of this observable would be very useful in constraining the region with low $m_{H^+}$ and large $\tan\beta$. Such results also provide complementarity to the different $B$ decays discussed above. The MSSM model dependence of the obtained limit is weak, as expected for a tree level observable. A similar exclusion region is obtained in the CMSSM.

\subsection{Muon $g-2$}
The anomalous magnetic moment of the muon $a_\mu = (g-2)/2$ receives non-zero contributions from radiative corrections. It has been determined to high precision both theoretically and experimentally, and can therefore be used to probe new physics effects, including the MSSM. 
The latest measured value for $a_\mu$ based on $e^+ e^-$ data is \cite{Bennett:2006fi}
\begin{equation}
a_\mu^\mathrm{exp} = (11\,659\,208.0\pm6.3)\times 10^{-10}.
\end{equation} 
Comparing this precise measurement to the SM prediction \cite{Miller:2007kk}
\begin{equation}
a_\mu^\mathrm{SM} = (11\,659\,178.5\pm6.1)\times 10^{-10}
\end{equation} 
leads to the discrepancy 
\begin{equation}
\delta a_\mu=a_\mu^\mathrm{exp}-a_\mu^\mathrm{SM}=(29.5\pm8.8)\times 10^{-10},
\end{equation} 
corresponding to a $3.4\,\sigma$ deviation from the SM.
The 95\% C.L. allowed range, including uncertainties from two loop SUSY corrections which have not been included, is:
\begin{equation}
11.5  \times 10^{-10} < \delta a_\mu < 47.5 \times 10^{-10}.
\label{eq:amu}
\end{equation}
\begin{figure}[!t]
\begin{center}
\includegraphics[width=7cm,keepaspectratio]{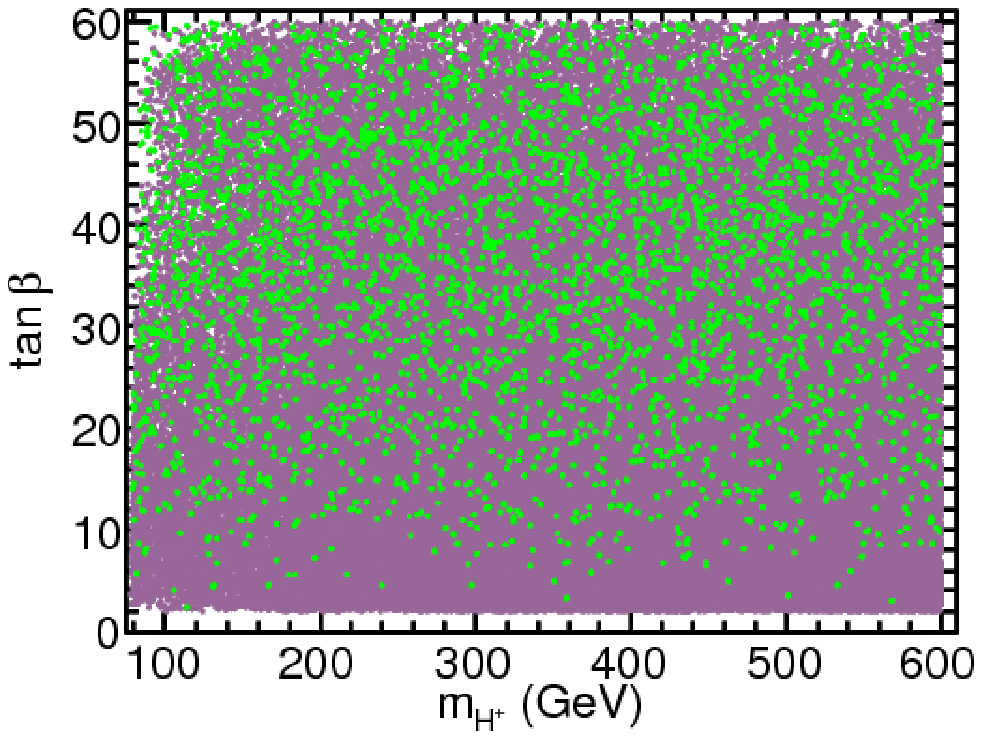}\hspace*{1cm}
\includegraphics[width=7cm,keepaspectratio]{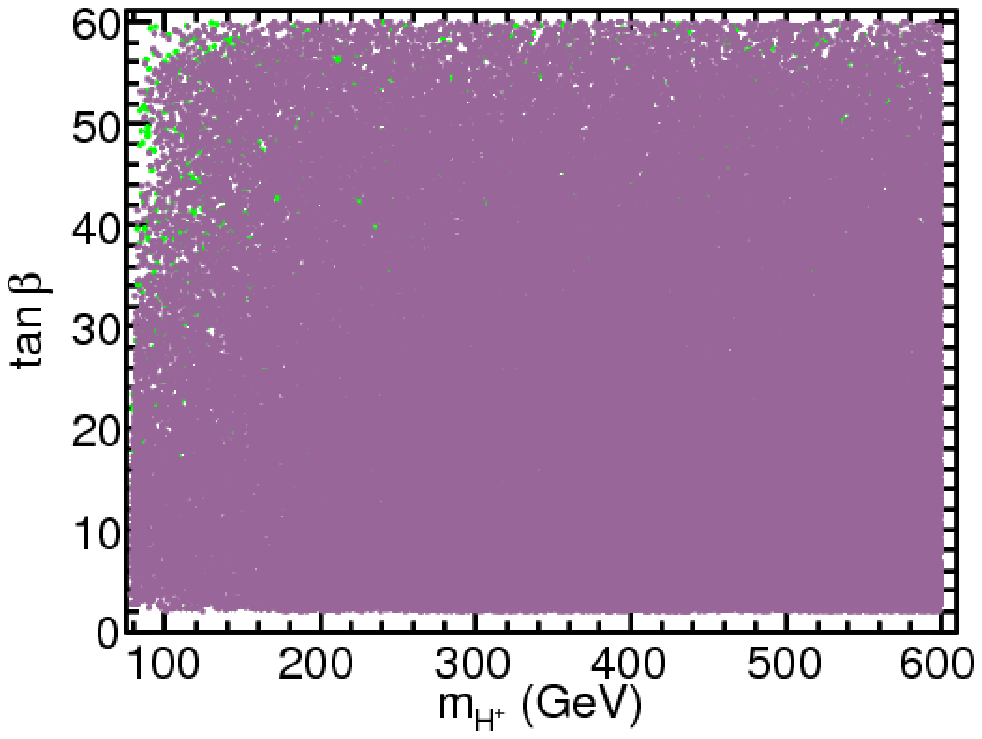}
\end{center}
\caption{Constraints from the muon anomalous magnetic moment $\delta a_\mu$ on the six dimensional NUHM parameter space, projected onto the plane $(m_{H^+},\tan\beta)$. The left plot shows the allowed points (green) in the foreground, whereas the right plot shows instead the excluded points (purple) in the foreground.}
\label{fig:damu}
\end{figure}%
In the MSSM, the discrepancy in $\delta a_\mu$ can be accounted for by the contributions from loops with exchange of neutralinos--smuons and charginos--sneutrinos. The result of imposing the constraint (\ref{eq:amu}) to the NUHM points is presented in Figure~\ref{fig:damu}, which shows no correlation between $\delta a_\mu$ and $(m_{H^+},\tan\beta)$. For the CMSSM, the $\delta a_\mu$ constraint acts differently, and such a correlation is observed.

\begin{figure}[!t]
\begin{center}
\includegraphics[width=7cm,keepaspectratio]{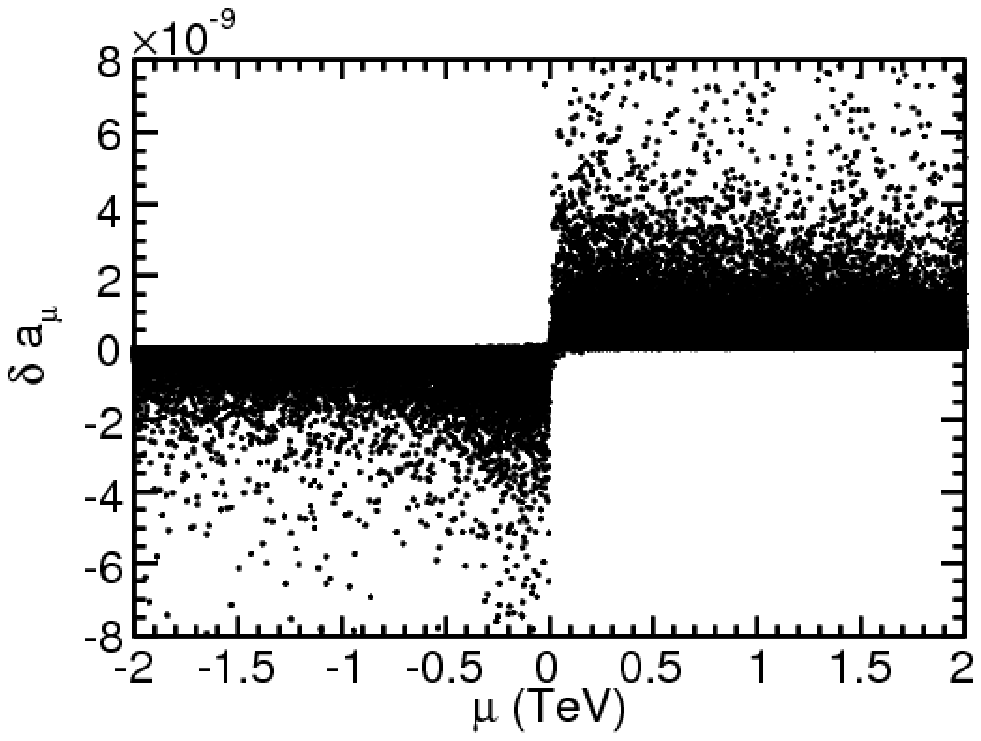}\hspace*{1cm}
\includegraphics[width=7cm,keepaspectratio]{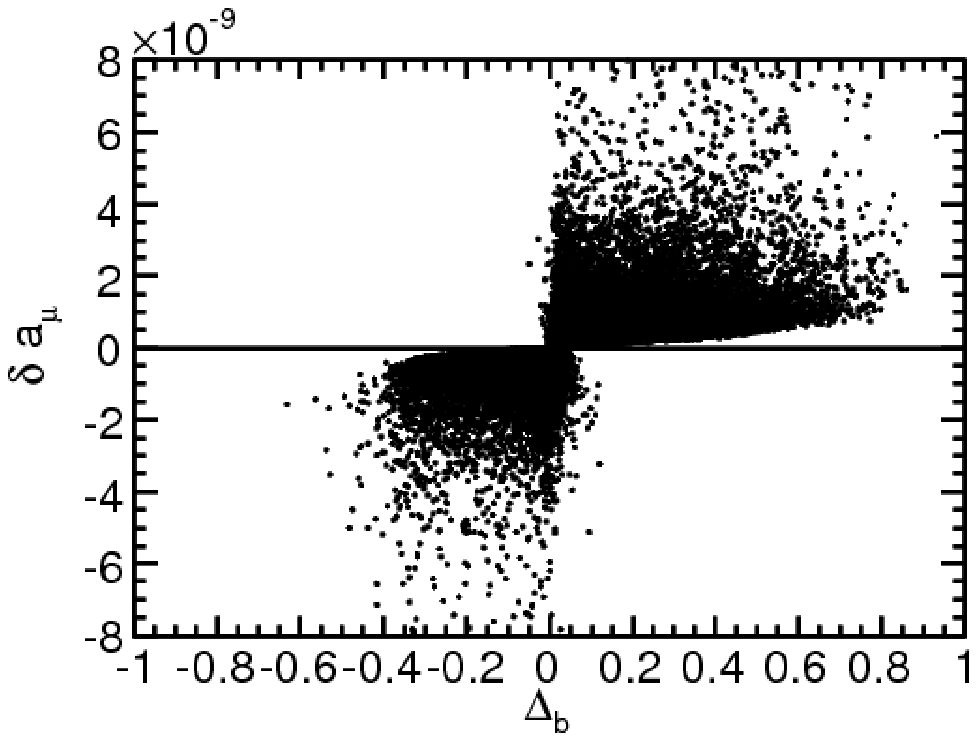}
\end{center}
\caption{Dependence on the $\mu$ parameter of $\delta a_\mu$ (left). Correlations between contribution to $\delta a_\mu$ and $\Delta_b$ (right).}
\label{fig:damu2}
\end{figure}

It is well-known that the sign of the MSSM contributions to $\delta a_\mu$ is directly coupled to the sign of the $\mu$ parameter. We illustrate this fact for the NUHM models in Figure~\ref{fig:damu2}. In the following, we will therefore take the constraint (\ref{eq:amu}) as a requirement of positive $\mu$ values. For a recent and more detailed discussion of this issue, see \cite{Feroz:2008wr}.

Since the sign of the leading contribution to $\Delta_b$ is determined by $\mu$ [compare Eq.~(\ref{eq:eps0})], the $\delta a_\mu$ results can also be used to constrain the favored range for these corrections. This is illustrated for the NUHM points in Figure~\ref{fig:damu2}. If the constraint $\mu>0$ is taken as an \emph{a priori} requirement on the MSSM model, it means a reduced model sensitivity for all charged Higgs observables at high $\tan\beta$.

\subsection{Dark Matter density}
When neutral, the lightest supersymmetric particle (LSP) is a suitable candidate for the cold dark matter of the universe. In the MSSM, the dark matter is therefore usually expected to consist of the lightest neutralino. The latest 5-year WMAP data \cite{Komatsu:2008hk} provides a precise experimental determination of the dark matter density:
\begin{equation}
\Omega_{\mathrm{DM}} h^2 = 0.1143\pm 0.0034,
\end{equation}
which can be compared to the theoretical calculation of the LSP relic density. Assigning a residual $10\%$ theoretical error \cite{Baro:2007em} in the prediction, we obtain the allowed interval at 95\% C.L.
\begin{equation}
0.094 < \Omega_{\mathrm{DM}} h^2 < 0.135 .
\end{equation}
However, as was shown in \cite{Arbey:2008kv}, it is not safe to use the lower bound in the above interval due to cosmological uncertainties in the era prior to Big Bang Nucleosynthesis. Also, dark matter can be composed of different components in addition to the LSP, which would again falsify the lower limit. We therefore discard the lower bound and use only
\begin{equation}
\Omega_{\mathrm{DM}} h^2 < 0.135
\label{eq:omega}
\end{equation}
to extract constraints from the relic density. 

\begin{figure}[!t]
\begin{center}
\includegraphics[width=7cm,keepaspectratio]{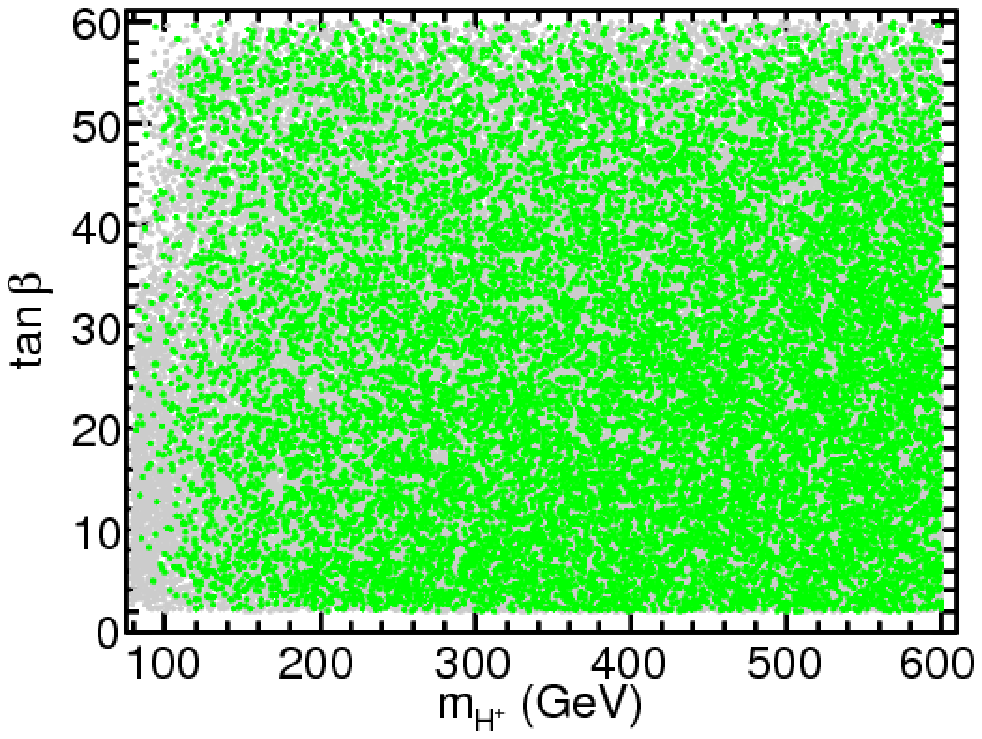}\hspace*{1cm}
\includegraphics[width=7cm,keepaspectratio]{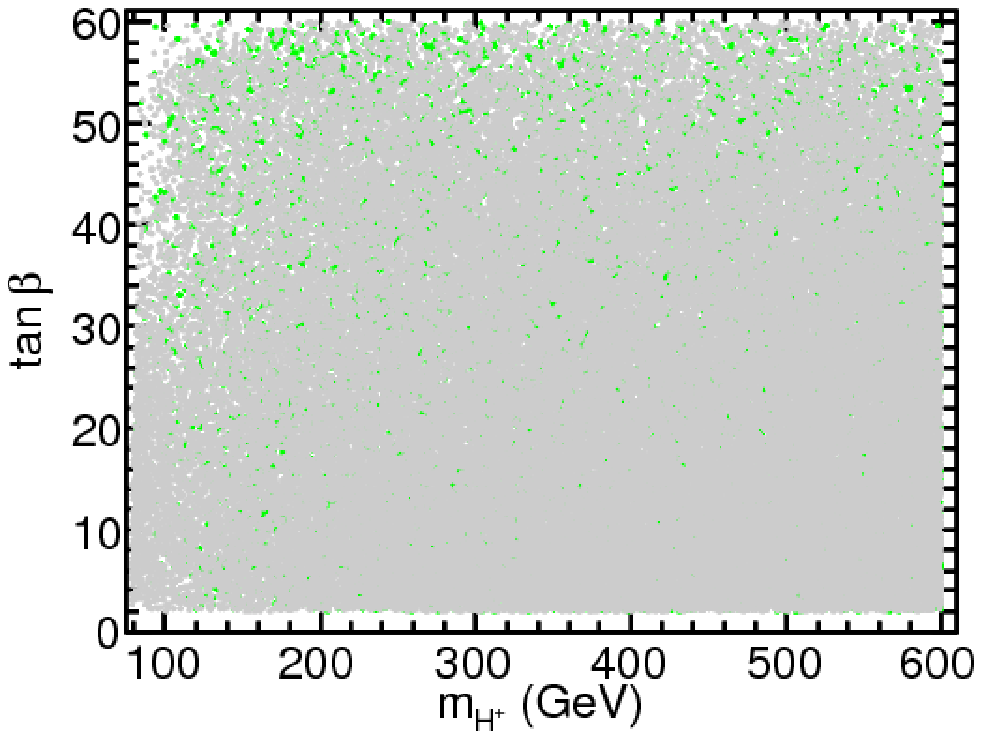}
\end{center}
\caption{Constraints from the density of dark matter on the six dimensional NUHM parameter space, projected onto the plane $(m_{H^+},\tan\beta)$. The left plot shows the allowed points (green) in the foreground, whereas the right plot shows instead the excluded points (light grey) in the foreground.}
\label{fig:omega}
\end{figure}

Requiring a neutral LSP, and that the constraint (\ref{eq:omega}) is satisfied, Figure~\ref{fig:omega} shows the results on the $(m_{H^+},\tan\beta)$ plane. The conclusion we draw from this figure is that the cosmological constraints on the relic density do not lead to distinct constrained ranges in $(m_{H^+},\tan\beta)$. This is also the case in the CMSSM. 

\subsection{Combined constraints and limits}
\begin{figure}
\begin{centering}
\includegraphics[height=6.5cm,keepaspectratio]{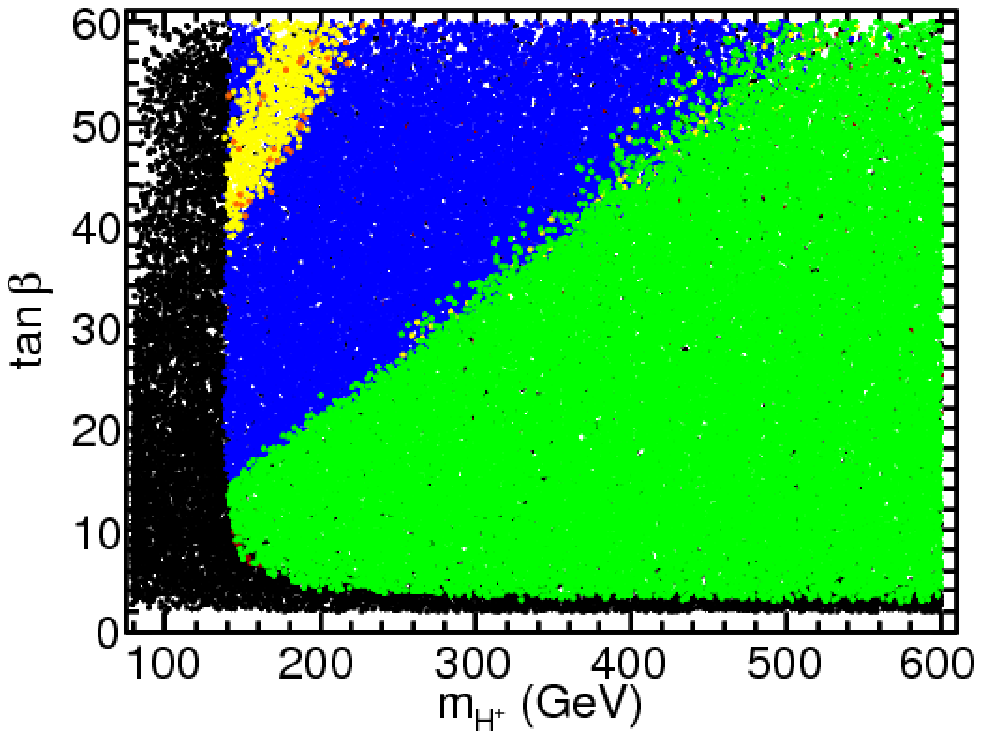}
\includegraphics[height=6.5cm]{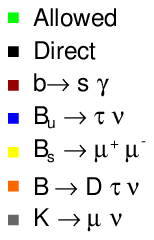}
\caption{Combined exclusion in NUHM models by different constraints, as described in the text. The constraints are applied in the order they appear in the legend, and the color coding corresponds to the first constraint by which a point is excluded. All points have $\mu>0$ and a neutral LSP.}
\label{fig:combined_NUHM}
\end{centering}
\end{figure}

In Figure~\ref{fig:combined_NUHM} we show a combination of constraints applied to the NUHM model points. The result is projected on the $(m_{H^+},\tan\beta)$ plane. The constraints are applied in the order indicated by the legend, and the first constraint by which a certain point is excluded determines its color. Points which are not excluded by any constraint are termed \emph{allowed} and displayed in the foreground to indicate the parameter regions still open for $H^+$.  Only points with $\mu>0$ and a neutral LSP are shown. From the figure, we note that the allowed points fall in a distinct region, forming a triangular shape in the lower half plane. The region of allowed points shares a diffuse boundary with that excluded by $B_u\to\tau\nu_\tau$ transitions. This diffuseness is the result of $\epsilon_0$ variations. 
\begin{figure}
\begin{centering}
\includegraphics[height=6.5cm]{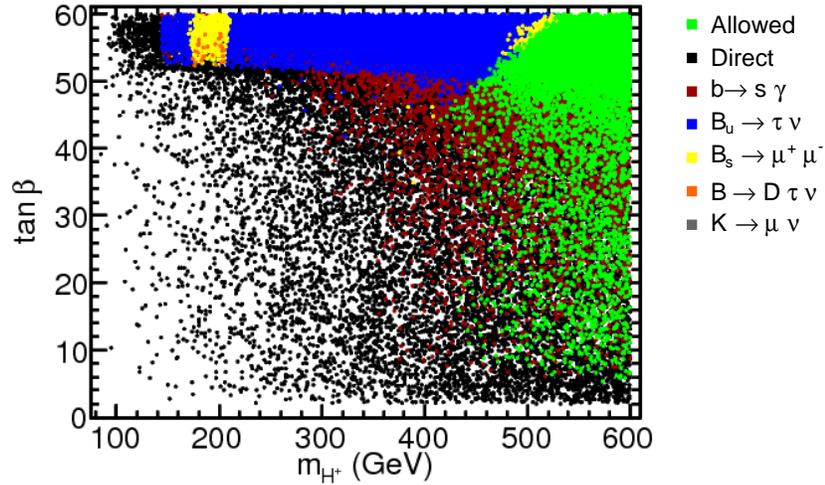}
\includegraphics[height=6.5cm]{legend}
\caption{Combined exclusion in CMSSM models by different constraints, as described in the text. The constraints are applied in the order they appear in the legend, and the color coding corresponds to the first constraint by which a point is excluded. All points have $\mu>0$ and a neutral LSP.}
\label{fig:combined_CMSSM}
\end{centering}
\end{figure}

Taking these constraints into account, we see that charged Higgs masses down to $m_{H^+}\simeq 135$~GeV can be accommodated, with the lowest masses allowed for intermediate $\tan\beta\sim 7$--$15$. For higher $\tan\beta$, the combined constraints follow the exclusion by $B_u\to\tau\nu_\tau$. The combined results are therefore particularly sensitive to the uncertainties associated with this decay channel. The allowed region is given approximately by $\tan\beta<(m_{H^+}/10$ GeV). Above the large region excluded by $B_u\to\tau\nu_\tau$ appears a smaller, mostly yellow, region containing points excluded by one or more of the constraints from $B_s\to\mu^+\mu^-$, $B\to D\tau\nu_\tau$, and $K\to\mu\nu_\mu$. Although the general indication is that $H^+$ is excluded in this region, a more conservative conclusion would be that it is still possible to find points excluded by only one of the three constraints. Since these can each be questioned on different grounds, there might be some room open for alternative interpretations.

As a final remark on the NUHM model, we note that the constraints from the well-established $b\to s\gamma$ transitions are not particularly strong in this class of models. The primary exclusion region for $b\to s\gamma$ at low $m_{H^+}$ has already been excluded by the direct constraints, and the remainder of the excluded points do not form a distinct exclusion region in the ($m_{H^+},\tan\beta$) plane.

Turning now to the CMSSM, we show in Figure~\ref{fig:combined_CMSSM} the same combination of constraints applied to this model. In the CMSSM, the mass scale for the Higgs bosons is not a free parameter, but it is fixed by the universality assumptions at the GUT scale through RGE running. Lower masses for the heavy MSSM Higgs bosons requires the $\tan\beta$ enhanced contribution from $y_b$ to cancel the always large RGE effects from the top Yukawa coupling $y_t$. The result of this balancing is seen in the figure, where the distribution of points reveals a clear preference for large $m_{H^+}$, and where smaller values for $m_{H^+}$ are only obtained in combination with high $\tan\beta\gtrsim 50$.

The combined constraints in the CMSSM work similarly as for the NUHM in limiting the parameter space available for the charged Higgs, with some important differences. The direct limits are very effective in ruling out parameter space regions with low--intermediate $\tan\beta$ and $m_{H^+}<400$ GeV. Constraints from $B_s\to\mu^+\mu^-$ exclude high $\tan\beta$ up to $m_{H^+}\simeq 500$ GeV. In the CMSSM in general $|\Delta_b|<0.5$ which reduces the MSSM model dependence. The same caveats discussed above apply when interpreting the constraints in the CMSSM for low $m_{H^+}$ and high $\tan\beta$. However, even with some of the more uncertain constraints removed, the allowed region in ($m_{H^+},\tan\beta$) still appears to be minimal. Nevertheless, this region around $m_{H^+}\simeq 200$ GeV is very interesting for early LHC running. 

\section{Charged Higgs at hadron colliders}
\label{sect:experiment}
There are two primary modes through which charged Higgs bosons can be produced at hadron colliders. When  the charged Higgs is light ($m_{H^+}<m_t-m_b$), it is kinematically accessible in the decay $t\to bH^+$ of the top quark. The decay width in the large $\tan\beta$ limit reads \cite{Carena:NPB:2000}
\begin{equation}
\Gamma(t\to bH^+)= \frac{g^2|V_{tb}|^2}{64\pi M_W^2}m_t\left(1-\frac{m^2_{H^+}}{m^2_t}\right)^2\frac{\overline{m}_b^2\tan^2\beta}{(1+\Delta_b)^2}\Bigl[1+\mathcal{O}(\alpha_s)\Bigr].
\label{eq:tbH}
\end{equation}
In favorable cases with small $m_{H^+}$ and large $\tan\beta$, the 2HDM branching ratio ($\Delta_b=0$) may reach values up to $\BR{t}{bH^+}\simeq 0.3$--$0.4$. The SUSY corrections entering through $\Delta_b$ can modify the pure 2HDM value substantially.

When the charged Higgs becomes too heavy to appear in the decay of on-shell top quarks, the favored mode of $H^+$ production is in association with a single top quark. For this mode of production, proper matching is required between the twin processes $gg$/$bg$ \cite{Alwall:2004xw}. The full NLO calculation is also available \cite{Zhu:2001nt,Plehn:2002vy,Berger:2003sm}, and we use a parametrization of this cross section for the comparison to experimental results below. The 2HDM cross section is then augmented with the appropriate $\tan\beta$ enhanced corrections proportional to $1/(1+\Delta_b)^2$.

The decay of the light $H^+$ proceeds mainly through one of the two channels $H^+\to\tau^+\nu_\tau$ which dominates for $\tan\beta\gtrsim 2$, or $H^+\to c\bar{s}$, becoming important for smaller $\tan\beta$ values. The $\tan\beta$ enhanced corrections to $H^+\to\tau^+\nu_\tau$ are negligible, since there are no SUSY-QCD corrections to the leptonic final state. Consequently, the width is given simply by
\begin{equation}
\Gamma(H^+\to\tau^+\nu_\tau)=\frac{g^2}{32\pi M_W^2}m_{H^+}m_\tau^2\tan^2\beta .
\end{equation}

For a heavier charged Higgs, the decay $H^+\to t\bar{b}$ opens up, and quickly overtakes $H^+\to\tau^+\nu_\tau$ as the dominant mode. In the large $\tan\beta$ regime, the partial width
\begin{equation}
\Gamma(H^+\to t\bar{b})=\frac{g^2|V_{tb}|^2N_c}{32\pi M_W^2}m_{H^+}\left(1-\frac{m_t^2}{m_{H^+}^2}\right)^2\frac{\overline{m}_b^2\tan^2\beta}{(1+\Delta_b)^2}\Bigl[1+\mathcal{O}(\alpha_s)\Bigr]
\label{eq:Htb}
\end{equation}
is proportional to $m_b^2$, and thus affected by SUSY corrections in the same way as $\Gamma(t\to bH^+)$. For all numerical evaluations of the $H^+$ branching ratios we use HDECAY \cite{Djouadi:1997yw}, which includes both QCD and MSSM corrections in a consistent fashion.

\subsection{SUSY decay modes}
\begin{figure}
\begin{centering}
\includegraphics[height=5.2cm]{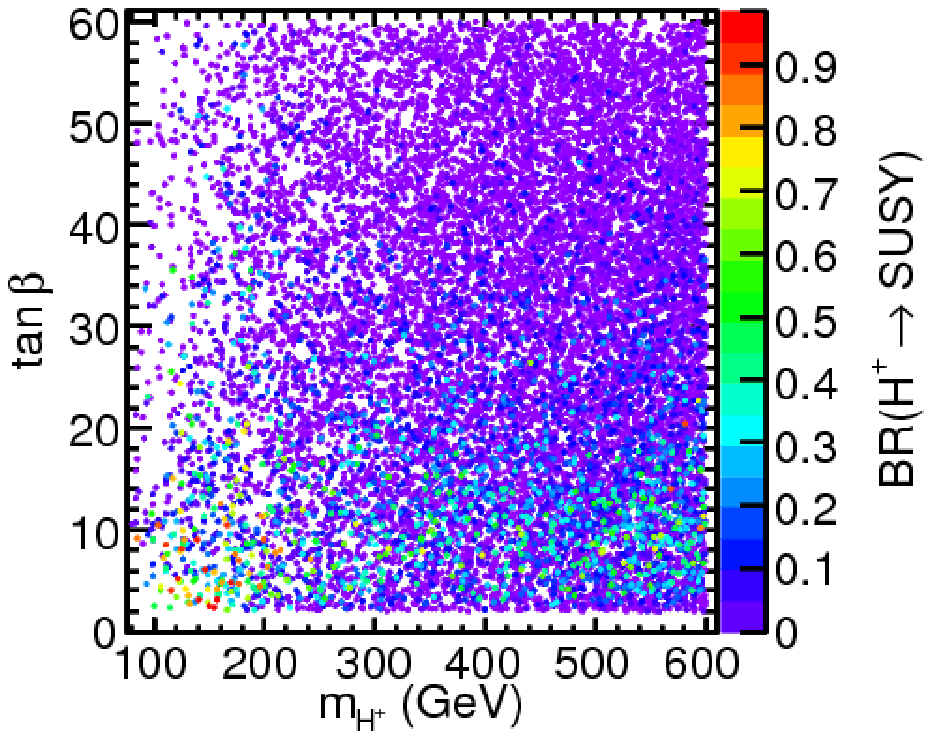}\hspace*{1cm}
\includegraphics[height=5.2cm]{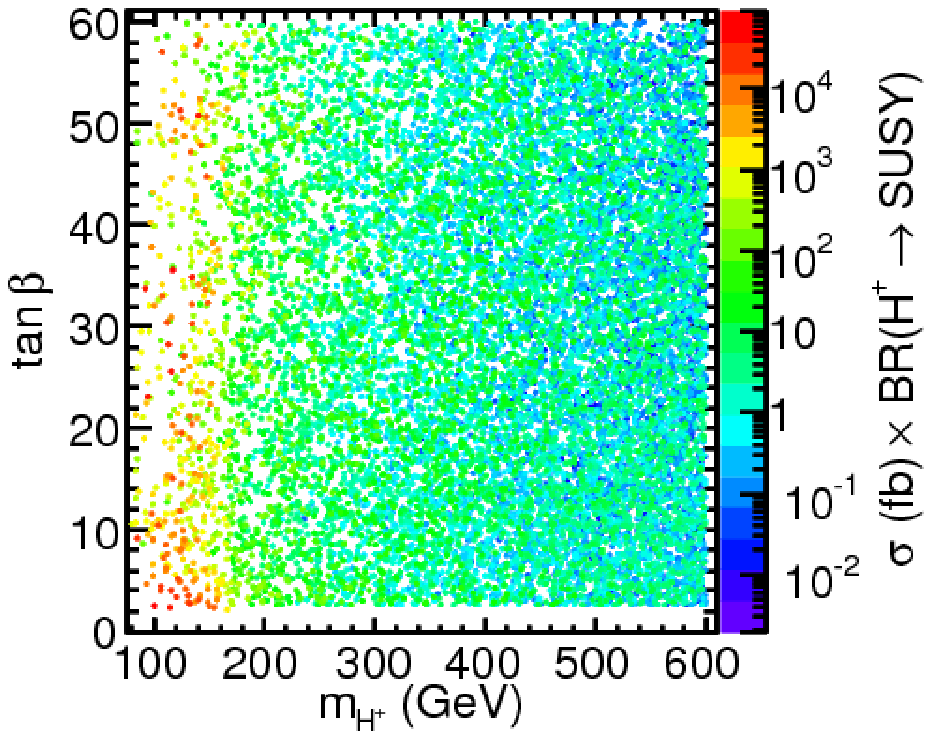}
\caption{Total branching ratio for $H^+$ into SUSY particles for the NUHM points, projected on the $(m_{H^+},\tan\beta)$ plane  (left). The branching ratio multiplied by the charged Higgs production cross section (right).}
\label{fig:SUSY1}
\end{centering}
\end{figure}

When allowed by the kinematics, the charged Higgs may decay to SUSY partners of the SM particles. Figure~\ref{fig:SUSY1} shows the total branching ratio for all SUSY decays of $H^+$ in the NUHM models. We see that the largest branching ratios are obtained in the intermediate region $\tan\beta=\sqrt{m_t/m_b}\sim 7$, where $\Gamma(H^+\to t\bar{b})$ has a minimum, and the detection of charged Higgs through the standard decay channels is most difficult.
On the other hand, in the right panel of Figure~\ref{fig:SUSY1} we show the branching ratio multiplied with the $H^+$ production cross section. This result comes out independent of $\tan\beta$.

\begin{figure}
\begin{centering}
\includegraphics[height=5.2cm]{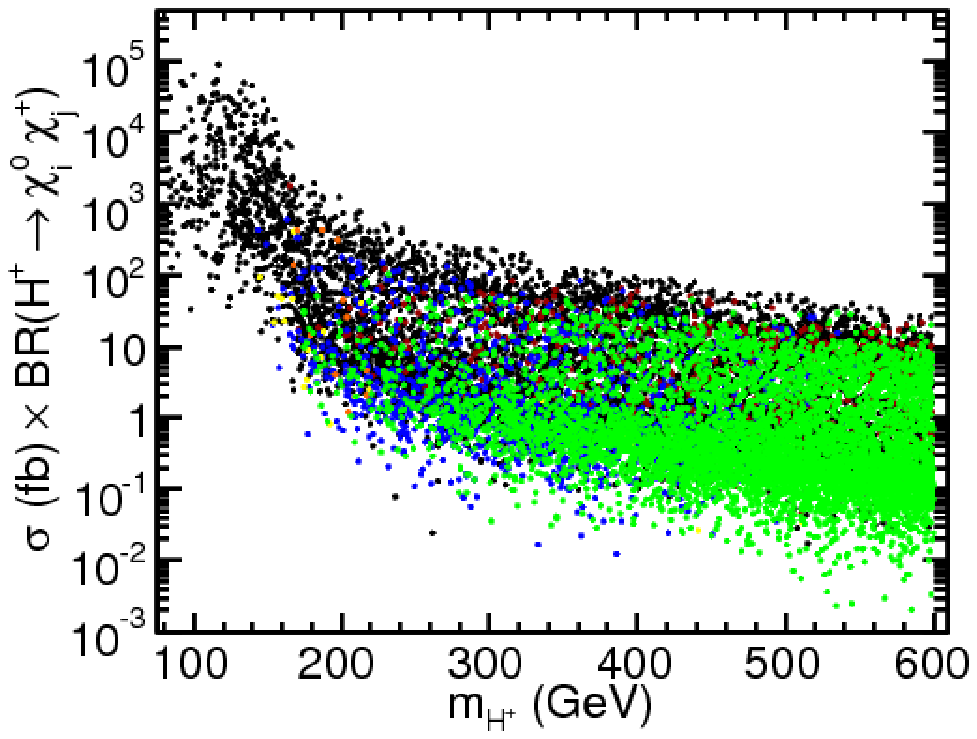}\hspace*{1cm}
\includegraphics[height=5.2cm]{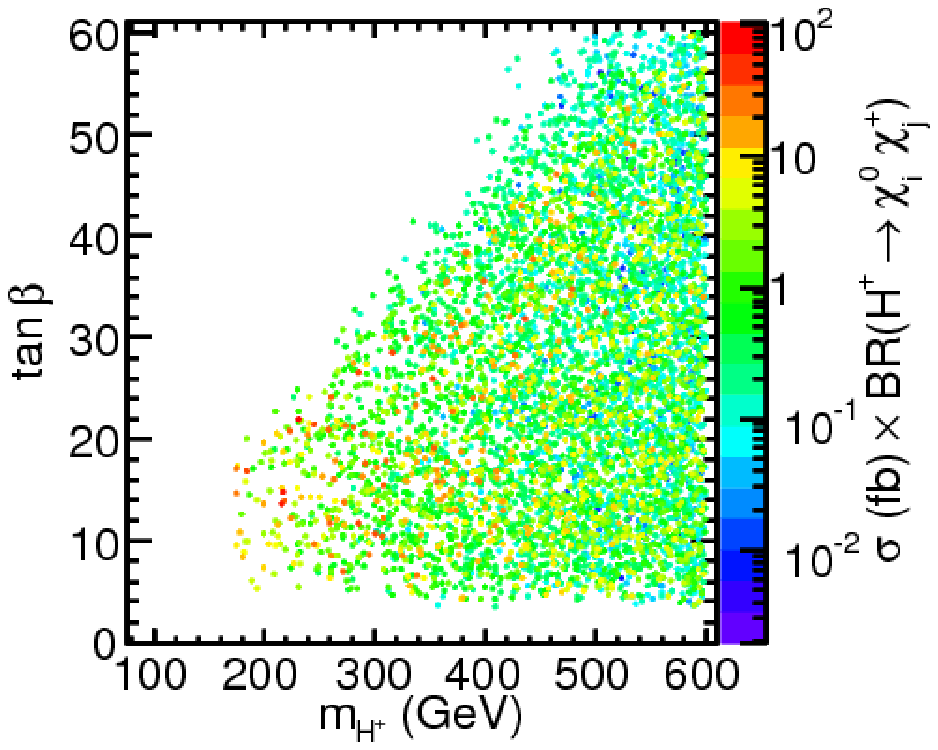}
\caption{Charged Higgs production cross section times $\BR{H^+}{\chi_i^0\chi_j^+}$, summed over all neutralino and chargino species. Direct and indirect constraint are applied as before. The color coding in the left plot agrees with that in Figure~11. In the right plot, only points allowed by the constraints are shown, and the color corresponds to $\sigma\times\rm{BR}$ (in fb). Both plots have requirements of $\mu>0$ and a neutral LSP.}
\label{fig:SUSY2}
\end{centering}
\end{figure}
We expect the main SUSY decay modes to be  $H^+\to \chi^+_i\chi^0_j$ \cite{Bisset:2000ud}. The experimental signatures for these modes depend on the further decay of the sfermions, where leptonic decays are likely to be required to suppress an otherwise overwhelming QCD background. For direct decay into the lightest chargino and the LSP, the final state may contain a single charged lepton and large missing $p_T$ from $\chi_1^+\to\chi_1^0 \ell^+\nu_\ell$. In the case where the charged Higgs decays to heavier charginos or neutralinos, the final state can become more involved. However, a promising generic signature is that based on three charged leptons and missing $p_T$ \cite{Bisset:2003ix}.

For the allowed points in the NUHM models we have verified that the chargino--neutralino decay modes are completely dominating, and that remaining SUSY decays can be neglected. Figure~\ref{fig:SUSY2} shows the sum of $\sigma\times\BR{H^+}{\chi_i^0\chi_j^+}$ for all chargino--neutralino channels together, with the constraints of section~\ref{sect:constraints} applied. We observe that in particular the direct mass constraints, and the constraints from $b\to s\gamma$, rule out points with low $m_{\chi_1^0}$ and $m_{\chi^+_1}$, which kinematically would give the highest number of events in the SUSY decay channels. Keeping in mind that Figure~\ref{fig:SUSY2} shows only the sum of all chargino--neutralino channels, and that no branching ratio into one lepton or three lepton final states has been applied, the total cross section is of the order of a few $100$ fb for the most promising parameter space points allowed by the constraints.

\subsection{Tevatron results}
At the Tevatron, CDF \cite{Abulencia:2005jd,CDF:2008} and D\O\ \cite{D0:2008} experiments have searched for light charged Higgs bosons in the decay of top quarks. The searches have been performed both in the $H^+\to\tau\nu_\tau$ and $H^+\to c\bar{s}$ channels , where the former is of course more interesting in the MSSM. The current best limit in the $H^+\to\tau^+\nu_\tau$ channel is obtained by D\O\ \cite{D0:2008} using $1$~fb$^{-1}$ of data. We show the model independent limit on $\BR{t}{H^+ b}$ from this search in Figure~\ref{fig:modind_NUHM}, assuming $H^+\to\tau^+\nu_\tau$ saturates the full width of $H^+$.

Recently, there has also emerged D\O\ results on a search for heavy charged Higgs in the $H^+\to t\bar{b}$ channel \cite{Abazov:2008rn}, but with limited sensitivity to the 2HDM (II) at this point.

\subsection{LHC prospects}
The kinematic range of the LHC will allow experiments to search both for light and heavy charged Higgs bosons. As discussed above, a heavy charged Higgs would preferentially decay through $H^+\to tb$. However, this channel has proven experimentally challenging. The decay mode of primary interest is therefore $H^+\to \tau^+\nu_\tau$ also when $m_{H^+}>m_t$, even though typically the $\BR{H^+}{\tau^+\nu_\tau}=10$--$15\%$ in the limit of high $m_{H^+}$. 

To determine the prospects for the LHC experiments to discover the charged Higgs boson in the MSSM models under study, we confront our model points with the experimental reach for a $5\,\sigma$ discovery obtained by ATLAS \cite{CSC} and CMS \cite{Baarmand&al:JPG:2006,Kinnunen:962050} through simulations. For both experiments, a full detector simulation is used, and systematic uncertainties are included. The discovery reach is reported for an integrated luminosity of $30$~fb$^{-1}$, corresponding to three years of LHC running at ``low luminosity''. 

\begin{figure}
\begin{centering}
\includegraphics[height=6.5cm,keepaspectratio]{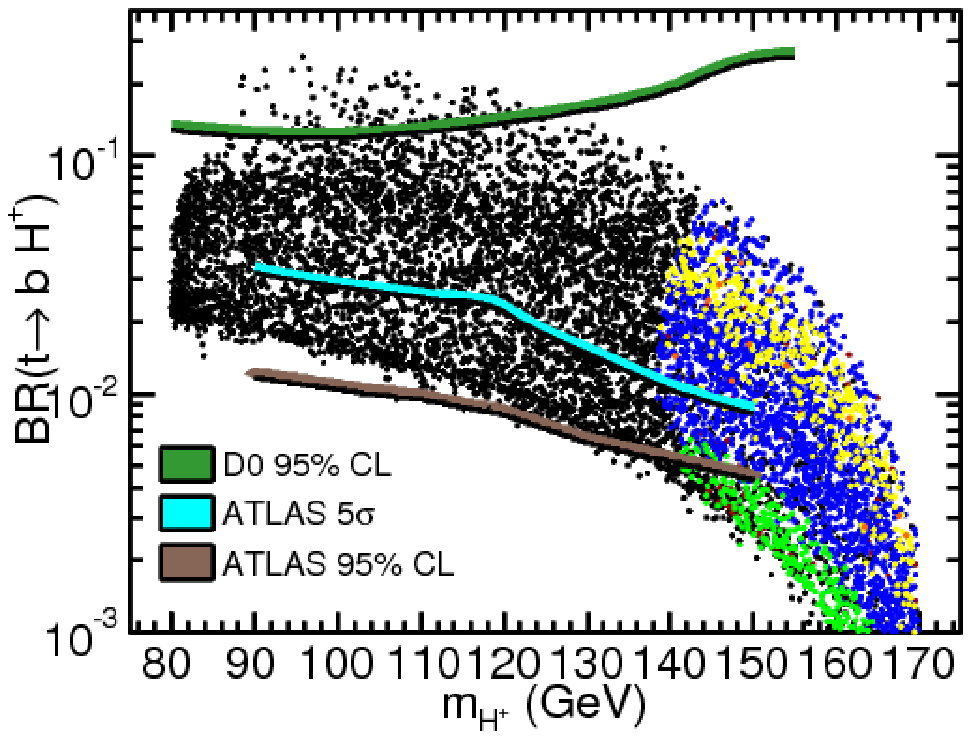}
\includegraphics[height=6.5cm]{legend}\vspace{0.3cm}
\includegraphics[height=6.5cm,keepaspectratio]{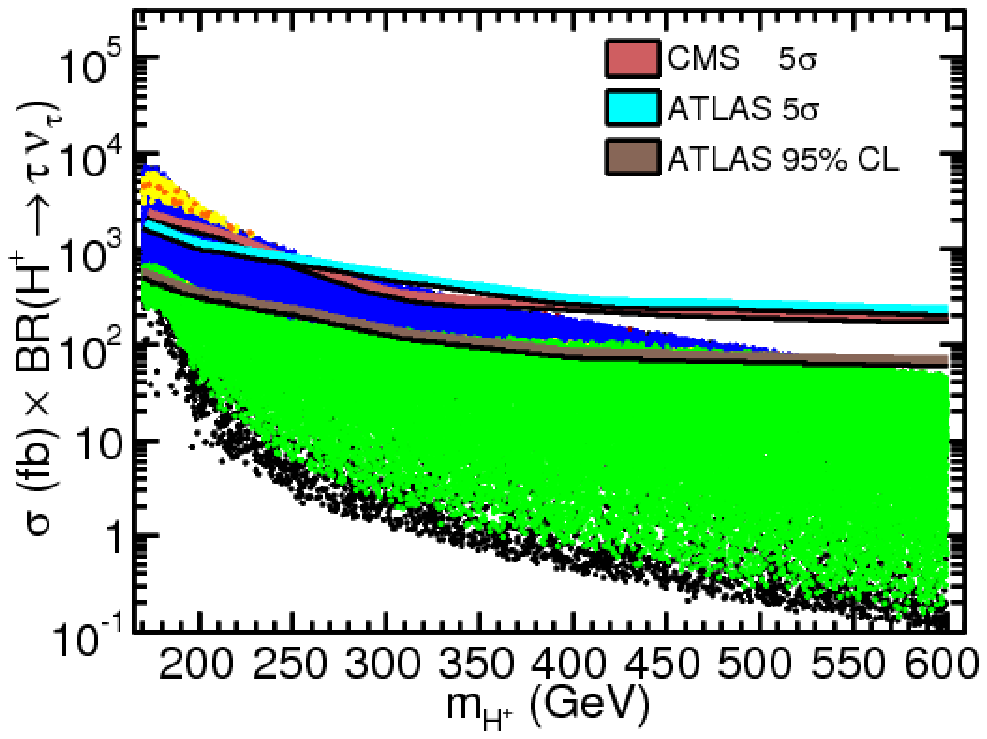}
\includegraphics[height=6.5cm]{legend}
\caption{Model-independent experimental discovery contours interleaved with NUHM model points. Color coding corresponds to exclusion by different constraints (see legend) which are explained in the text. Points allowed by the constraints (green) are always displayed in the foreground. A neutral LSP and $\mu>0$ is required.}
\label{fig:modind_NUHM}
\end{centering}
\end{figure}
The result for the NUHM model points is presented in Figure~\ref{fig:modind_NUHM}, showing the light and heavy $m_{H^+}$ cases separately. For the light $m_{H^+}$, the bulk of the NUHM points are accessible already with $30$ fb$^{-1}$. Since the points favored by the constraints lie close to the kinematic limit $m_{H^+}\to m_t-m_b$, dedicated studies are required to carefully evaluate the discovery prospects in this mass region. 
Figure~\ref{fig:modind_NUHM} reveals that the models excluded by the different constraints are also those which have the highest cross section and branching ratio for charged Higgs production. This is simply a result of the universal dependence on $\tan^2\beta/m^2_{H^+}$ shared by most $H^+$ observables in the high $\tan\beta$ limit. 

\begin{figure}
\begin{centering}
\includegraphics[height=6.5cm,keepaspectratio]{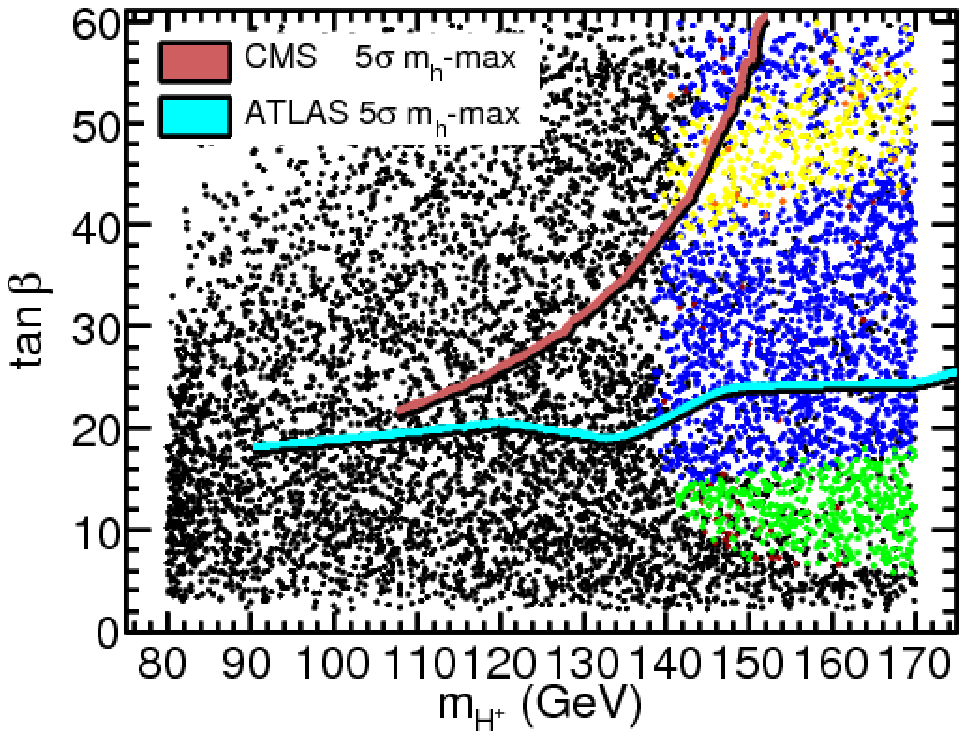}
\includegraphics[height=6.5cm]{legend}\vspace{0.3cm}
\includegraphics[height=6.5cm,keepaspectratio]{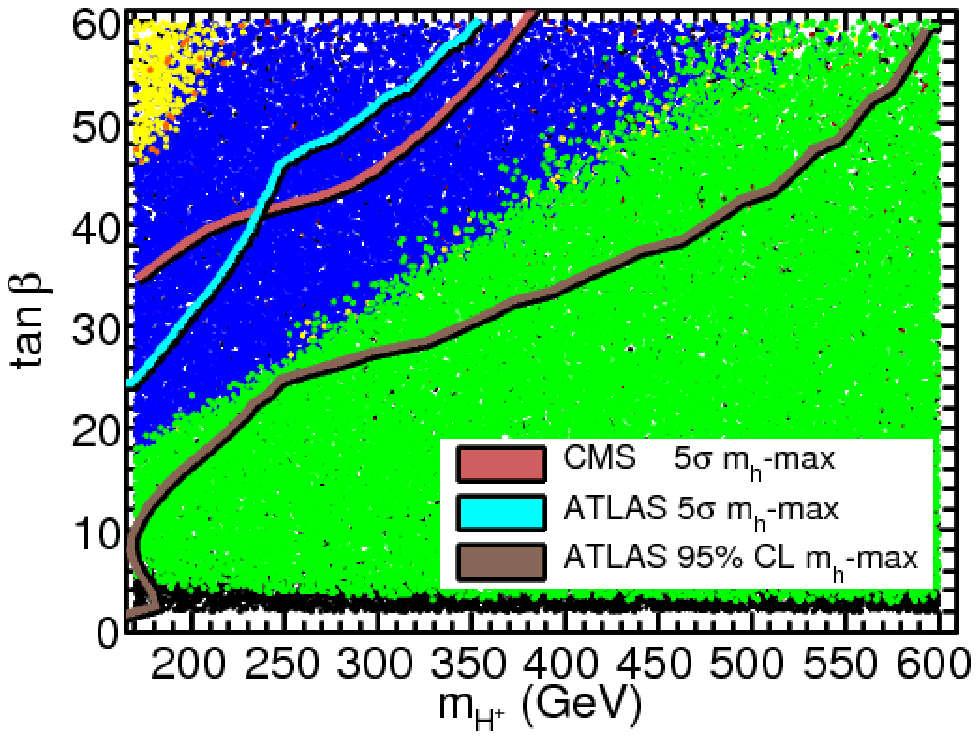}
\includegraphics[height=6.5cm]{legend}
\caption{Experimental discovery contours for the $m_h$--max scenario, interleaved with NUHM model points projected on the $(m_{H^+},\tan\beta)$ plane. Color coding corresponds to exclusion by different constraints (see legend) which are explained in the text. Points allowed by the constraints (green) are drawn in the foreground. A neutral LSP and $\mu>0$ is required. In the low mass case no ATLAS contour at $95\%$ CL is visible, since the reach for exclusion covers the whole plane.}
\label{fig:proj_NUHM}
\end{centering}
\end{figure}

Having compared the NUHM models to the experimental discovery reach in a model independent way, we now consider the interpretation in the $(m_{H^+},\tan\beta)$ plane. The experimental results \cite{CSC,Hashemi:2008ma} are presented in the $m_h$--max scenario, described in section~\ref{sect:susymodels}. A comparison between the experimental results and the NUHM points is given in Figure~\ref{fig:proj_NUHM}. The ATLAS results in this figure are obtained from a combined discovery contour, whereas the CMS results are reported as two contours for light and heavy $H^+$ separately. For ATLAS we also include a projected exclusion limit reported at the $2\,\sigma$ level \cite{CSC}. Figure~\ref{fig:proj_NUHM} illustrates even more explicitly than Figure~\ref{fig:modind_NUHM} the correspondence between the region with highest discovery reach for collider experiments and the most powerful exclusion by indirect constraints. 
As a side remark, it should be mentioned that $m_{H^+}<123$ GeV is already excluded in the $m_h$--max scenario by the direct limit on $m_h$ \cite{Schael:2006cr}. The use of $m_h$--max as a benchmark scenario in this region is therefore somewhat questionable.

\subsection{MSSM model dependence}
As a final point we discuss the sensitivity of the presented experimental results to the choice of MSSM benchmark scenario. This issue was recently discussed for CMS \cite{Hashemi:2008ma} in the context of $m_h$--max scenarios with different choices for the $\mu$ parameter. For any sub-dominant decay channel of a heavy charged Higgs, such as $H^+\to\tau^+\nu_\tau$, the effects of the bottom Yukawa corrections cancel to a large extent between the production and the decay. To see this, we consider the corrected cross section times the branching ratio
\begin{equation}
\sigma\times\BR{H^+}{\tau^+\nu_\tau}=\frac{\sigma_0}{(1+\Delta_b)^2}\frac{\Gamma_\tau}{\Gamma_\tau+\frac{\Gamma_{tb}}{(1+\Delta_b)^2}+\Gamma_\mathrm{X}}
\simeq\sigma_0\frac{\Gamma_\tau}{\Gamma}\left\{1-2\Delta_b\Bigl(1-\frac{\Gamma_{tb}}{\Gamma}\Bigr)\right\},
\label{eq:sigbr}
\end{equation}
where $\sigma_0$ is the cross section obtained in the pure 2HDM, $\Gamma_{tb}$ the uncorrected width for $H^+\to t\bar{b}$, and $\Gamma$ the uncorrected total width. $\Gamma_X$ refers to any decay mode which is not $\tau^+\nu_\tau$ or $tb$. Equation~(\ref{eq:sigbr}) shows that, when $\Gamma_\tau$ and $\Gamma_\mathrm{X}$ are both small with respect to $\Gamma_{tb}$, the combined $\Delta_b$ correction in this channel is second order in $\BR{H^+}{\tau^+\nu_\tau}$, therefore typically less than $10$--$15\%$ even for large values of $|\Delta_b|$.

To assess the model sensitivity of the ATLAS results presented in \cite{CSC}, we evaluate whether the NUHM models could lead to a $5\,\sigma$ discovery of the charged Higgs boson. For each model point, we determine either $\BR{t}{bH^+}$ or $\sigma(pp\to tH^+)$ as appropriate for the value of $m_{H^+}$, followed by the $\BR{H^+}{\tau^+\nu_\tau}$. This is done both with the MSSM $\Delta_b$ corrections applied, and for a fixed $\Delta_b=0$ corresponding to a pure 2HDM (II) with the same values for $(m_{H^+},\tan\beta)$.
\begin{figure}[!t]
\begin{center}
\includegraphics[width=7cm,keepaspectratio]{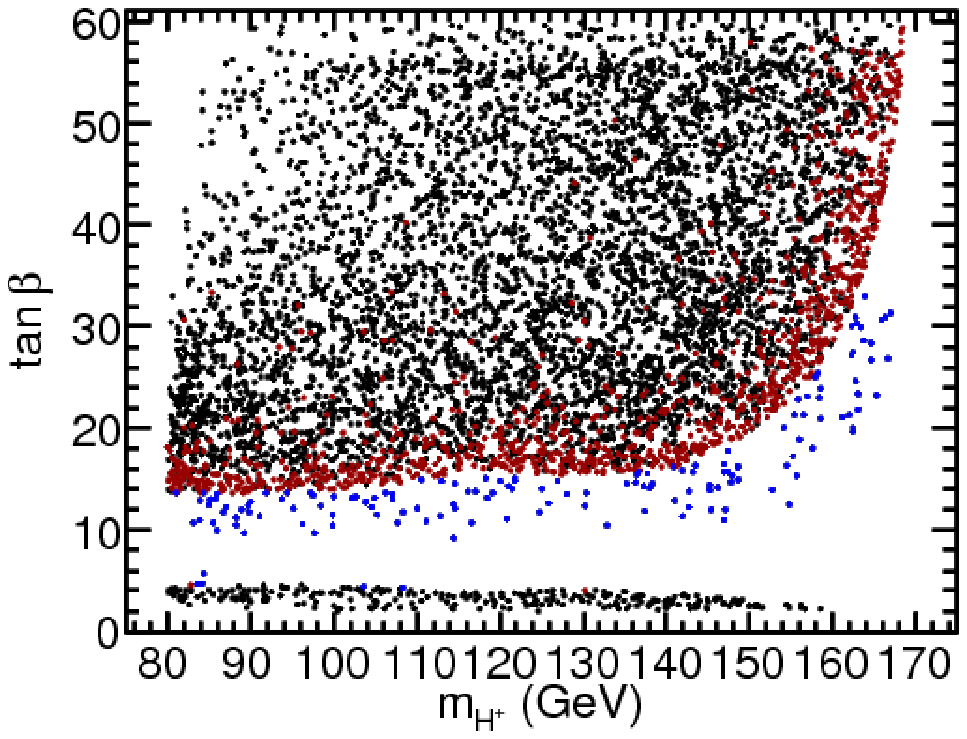}\hspace*{1cm}
\includegraphics[width=7cm,keepaspectratio]{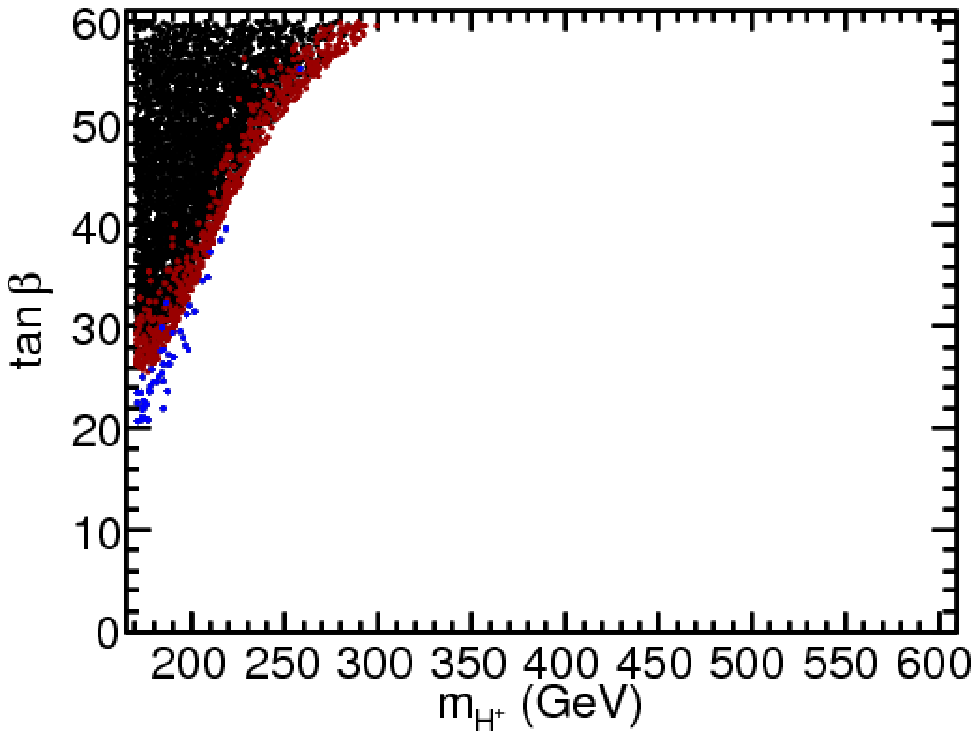}
\end{center}
\caption{Sensitivity of ATLAS discovery potential to $\Delta_b$ corrections. Points correspond to NUHM models which can be discovered (black), which can be discovered only because of $\Delta_b<0$ corrections (blue), and points which cannot be discovered because of $\Delta_b>0$ corrections (red). Both positive and negative values for $\mu$ are considered.}
\label{fig:expsens}
\end{figure}
We then compare the two numbers obtained to what is required for charged Higgs discovery with $30$ fb$^{-1}$. In Figure~\ref{fig:expsens} we show the result of this comparison. Points for which the cross section (branching ratio) is large enough for a $5\,\sigma$ discovery with the standard $\Delta_b$ corrections included are shown in black.\footnote[1]{Since we extrapolate the discovery contour in Figure~\ref{fig:modind_NUHM} above $m_{H^+}=150$ GeV, the distribution in Figure~\ref{fig:expsens} of points allowing a $5\,\sigma$ discovery does not exactly follow the ATLAS contour of Figure~\ref{fig:proj_NUHM} in this mass region. This visual difference is of no importance to our conclusions.} The subset of black points which would not be discovered with $\Delta_b=0$ are shown in blue. Finally, points which are hidden from discovery because of the $\Delta_b$ correction are shown in red. The red points are such that they would be accessible with $\Delta_b=0$.

As seen from the left plot in Figure~\ref{fig:expsens}, the $\BR{t}{bH^+}$ can be altered quite significantly by $\Delta_b$ corrections, resulting in a pronounced MSSM model dependence. There exist (red) points for such high values as $\tan\beta=50$ which do not allow a $5\,\sigma$ charged Higgs discovery. These points have large values of $\Delta_b\simeq 1$, thus correspond to large and positive $\mu$. In the NUHM models at high $\tan\beta$, the resulting distribution of theoretically allowed points is not uniform in $\mu$, but has a bias towards positive values. This model effect explains the dominance in number of red over blue points in Figure~\ref{fig:expsens}.

For a heavy charged Higgs the conclusion is different, as seen from the right plot of Figure~\ref{fig:expsens}. As expected from Eq.~(\ref{eq:sigbr}), the dependence on $\Delta_b$ is much milder for this case, resulting in a more concentrated distribution of red points. The experimental discovery contour is therefore fairly stable with respect to $\Delta_b$ variations. The absence of blue points at high $\tan\beta$ again results from the model preference for positive $\mu$ in this region.

\section{Summary and Conclusions}
\label{sect:conclusions}
Charged Higgs bosons are of special interest since they can provide definite signatures for physics beyond the Standard Model. In this paper we have analyzed the charged Higgs phenomenology in the constrained MSSM and in models with non-universal Higgs masses.

To investigate the experimental prospects for charged Higgs discovery, we have first examined direct and indirect constraints from a fairly complete set of flavor physics observables, calculated with the publicly available program SuperIso \cite{SuperIso}. We used results from $b \to s \gamma$, $B_u \to \tau \nu_\tau$, $B \to D \tau \nu_\tau$, $B_s \to \mu^+ \mu^-$, and $K \to \mu \nu_\mu$ transitions, together with the muon $(g-2)$ and cosmological constraints. In this manner we have identified the allowed regions for $m_{H^+}$ and $\tan\beta$ at $95\%$ C.L. The combined constraints from the flavor observables exclude the region with low $m_{H^+}$ and large $\tan\beta$. In the CMSSM, the lowest allowed value for $m_{H^+}$ is found to be of the order $m_{H^+}\gtrsim 400$ GeV, while $m_{H^+}\simeq 135$ GeV is still not excluded in the NUHM models. Even lower values could be obtained in the intermediate $\tan\beta$ region by relaxing the SM bound on $m_h$.

It is important not to over-interpret the limits obtained using indirect observables. We have shown explicitly that the results from $B_u \to \tau \nu_\tau$ in particular are subject to large uncertainties from the determination of $|V_{ub}|$. Likewise, the results obtained from $K \to \mu \nu_\mu$ are highly dependent on the value of $f_K/f_\pi$ from lattice QCD. Improvements in the measurements of $B$ physics observables, especially the $B_u \to \tau \nu_\tau$ and $B \to D \tau \nu_\tau$ branching ratios, would certainly be welcome and serve to refine the situation.

We compared the MSSM models, with the constraints applied, to the projected experimental sensitivities of ATLAS and CMS in the main charged Higgs discovery channels. This comparison illustrates that most of the indirect constraints are relevant in the same parameter space regions where the charged Higgs production cross section at the LHC is the largest. We have also considered the interpretation of the discovery potential in specific NUHM models. For $m_{H^+}<m_t$, we find a possibly large sensitivity to the MSSM benchmark scenario through corrections to the bottom Yukawa coupling, while this effect cancels to a large degree for the channel $H^+\to\tau^+\nu_\tau$ when $m_{H^+}>m_t$.

This study can easily be extended to scenarios with other mechanisms than gravity mediated supersymmetry breaking. More interesting would be to consider the MSSM beyond minimal flavor violation, including effects of $\cp$- and $\mathcal{R}$-parity violation, or to carry over the constraints on charged Higgs bosons to models with enlarged Higgs sectors like the NMSSM \cite{Akeroyd:2007yj}. The same observables discussed here, in particular those which involve tree-level exchange of the charged Higgs, play an important role in constraining any MSSM-like model. We therefore propose that complementary discovery channels, governed by couplings which are not constrained at this point, should be investigated to determine the prospects for early charged Higgs discovery at the LHC. Such a discovery would serve as an indication of a non-minimal model being realized in nature.

\section*{Acknowledgments}
We thank Gunnar Ingelman and Johan Rathsman for reading the manuscript, and for their useful comments and suggestions. We are also grateful to Martin Flechl and Elias Coniavitis for interesting discussions on charged Higgs boson searches.
\bibliographystyle{JHEP}
\bibliography{chsusy}

\end{document}